\documentclass[11pt,a4paper]{article}

\usepackage[T1]{fontenc}
\usepackage[utf8]{inputenc}
\usepackage{booktabs,circledsteps}

\usepackage{epigraph}
\makeatletter
\renewcommand{\epigraph}[2]{%
  \vspace{1ex}%
  \noindent\hfill
  \begin{minipage}{\epigraphwidth}
  \small
    \noindent #1\par
    \vspace{0.5ex}%
    \noindent #2%
  \end{minipage}%
  \vspace{0ex}%
}
\makeatother
\usepackage[margin={0.75in},centering]{geometry}
\usepackage{mathpazo}     
\usepackage{amsmath,bm,amssymb,amsthm}     
\usepackage{microtype}    
\usepackage{enumitem}  

\usepackage{tikz}
\usetikzlibrary{calc, trees, positioning, arrows}

\newtheorem{proposition}{Proposition}
\newtheorem{remark}{Remark}
\newtheorem{corollary}{Corollary}
\newtheorem{theorem}{Theorem}

\newtheorem{claim}{Claim}

\theoremstyle{definition}

\newtheorem{Example}{Example}

\newenvironment{example}
{\begin{Example}}
	{\hfill \(\diamondsuit\)\end{Example}}

\newcommand{\footautoref}[1]{\hyperref[#1]{Footnote~\ref*{#1}}}

\DeclareMathOperator*{\supp}{supp}
\DeclareMathOperator*{\marg}{marg}

\DeclareMathOperator*{\argmax}{arg\,max}

\newcommand{\rat}[1][]{%
  \ensuremath{%
    R
  }%
}

\makeatletter
\newcommand{\lk}[1][]{%
  \@ifnextchar^%
    {\lk@sup{#1}}%
    {\ensuremath{%
       L%
       \if\relax\detokenize{#1}\relax
       \else
         _{#1}%
       \fi
       [p]%
     }%
   }%
}

\def\lk@sup#1^#2{%
  \ensuremath{%
    L%
    \if\relax\detokenize{#1}\relax
    \else
      _{#1}%
    \fi
    ^{#2}[p]%
  }%
}
\makeatother

\usepackage{graphicx}
\usepackage{longtable,booktabs,tabularx}
\usepackage[flushleft]{threeparttable}
\usepackage{multirow,array,caption}
\usepackage{pdflscape}      
\usepackage{color,colortbl} 
\usepackage{subcaption}

\definecolor{darkblue}{rgb}{0.15,0,0.37}
\definecolor{darkred}{rgb}{0.35,0,0.08}
\definecolor{mygrey}{rgb}{0.85,0.85,0.85}

\newcommand*{\myalign}[2]{\multicolumn{1}{#1}{\(#2\)}}

\usepackage[authoryear]{natbib}
\usepackage[
pagebackref=false,
breaklinks=true,
colorlinks=true,
linkcolor=black,
citecolor=black,
urlcolor=black,
pdfstartview=FitV,
unicode=true
]{hyperref}



\usepackage[onehalfspacing]{setspace} 
\usepackage[marginal]{footmisc}       
\usepackage{fancyhdr}
\usepackage{lastpage}

\newcommand{\ShortTitle}[1]{\def\theShortTitle{#1}}
\newcommand{\ShortAuthors}[1]{\def\theShortAuthors{#1}}

\newcommand\blfootnote[1]{%
	\begingroup
	\renewcommand\thefootnote{}\footnote{\hspace*{0pt}#1}%
	\addtocounter{footnote}{-1}%
	\endgroup
}

\makeatletter
\let\orig@makefntext\@makefntext

\renewcommand{\@makefntext}[1]{\parindent 0pt \noindent #1}
\makeatother

\newenvironment{niceabstract}
{   \vspace{-2.85em}
	\begin{quote}
    \normalsize
		\setlength{\parskip}{0em}
		\setlength{\parindent}{0em}
	}
	{
	\end{quote}
}

\newcommand{\PaperTitle}[1]{\def\thePaperTitle{#1}}

\newcommand{\AuthorOne}[5]{%
	\def\authorOneFirstName{#1}%
	\def\authorOneLastName{#2}%
	\def\authorOneAff{#3}%
	\def\authorOneLoc{#4}%
	\def\authorOneEmail{#5}%
}

\newcommand{\AuthorTwo}[5]{%
	\def\authorTwoFirstName{#1}%
	\def\authorTwoLastName{#2}%
	\def\authorTwoAff{#3}%
	\def\authorTwoLoc{#4}%
	\def\authorTwoEmail{#5}%
}

\newcommand{\AutoShortAuthors}{%
	\ShortAuthors{\authorOneLastName\ and~\authorTwoLastName}%
}

\newcommand{\printTitleAndAuthors}{
	\thispagestyle{empty}
	\begin{center}
		{\Large\thePaperTitle}\\[1em]
	\end{center}
	
	\noindent
	\begin{minipage}[t]{0.45\textwidth}
		\centering
		{\normalsize \authorOneFirstName\ \authorOneLastName}\\[-0.5em]
		\vspace{4pt}
		{\small
			\authorOneAff \\[-0.33em]
			\authorOneLoc \\[-0.33em]
			\href{mailto:\authorOneEmail}{\texttt{\authorOneEmail}}
		}
	\end{minipage}
	\hfill
	\begin{minipage}[t]{0.45\textwidth}
		\centering
		{\normalsize \authorTwoFirstName\ \authorTwoLastName}\\[-0.5em]
		\vspace{4pt}
		{\small
			\authorTwoAff \\[-0.33em]
			\authorTwoLoc \\[-0.33em]
			\href{mailto:\authorTwoEmail}{\texttt{\authorTwoEmail}}
		}
	\end{minipage}
	
	\vspace{0.5em}
	\begin{center}
		{\normalsize \today}
	\end{center}
}

\begin{document}
	
	\ShortTitle{Reasoning About Bounded Reasoning}
	
	\PaperTitle{Reasoning About Bounded Reasoning}
	
	\AuthorOne{Shuige}{Liu}
	{Department of Decision Sciences \\[-0.33em] Bocconi University}
	{Milan, Italy}
	{shuige.liu@unibocconi.it}
	
	\AuthorTwo{Gabriel}{Ziegler}
	{Freie Universit\"at Berlin and \\[-0.33em] Berlin School of Economics}
	{Berlin, Germany}
	{gabriel.ziegler@fu-berlin.de}
	
	\AutoShortAuthors
	
	\printTitleAndAuthors
	
	\blfootnote{Earlier versions of some of our results were circulated under the title ``Level-\(k\) Reasoning, Cognitive Hierarchy, and Rationalizability'' by \citet{sliudraft01}. Special thanks for detailed and very valuable comments go to Pierpaolo Battigalli, Amanda Friedenberg, Terri Kneeland, and Antonio Penta. Furthermore, we gratefully acknowledge helpful comments and suggestions from Matteo Bizzarri, Meng-Jhang Fong, Elshan Garashli, Michael Greinecker, Pierfrancesco Guarino, Steven Kivinen, Christoph Kuzmics, Fabio Maccheroni, Niccol\`o Lomys, Marco Pagnozzi, Thi Quynh Lien Pham, Mikl\'{o}s Pint\'{e}r, Marciano Siniscalchi, and audience members at the QSMS seminar at the Budapest University of Technology and Economics, the 3rd Italian Junior Workshop on Economic Theory at the University of Naples Federico II, the York-Durham-Edinburgh Theory Workshop 2025, the 24th Annual SAET Conference in Ischia, Italy, the 3rd Workshop on Experimetrics \& Behavioral Economics in Soleto, Italy, the XIX GRASS Workshop in Padova, Italy, the 17th Nordic Conference in Behavioral and Experimental Economics at the University of Stavanger, Norway, the Berlin School of Economics Microeconomic Colloquium, the seminar at the University of Reading, and the Roy seminar at the Paris School of Economics. All remaining errors are ours.}
	
	\begin{niceabstract}
\textbf{\normalsize Abstract.}
In experimental applications of bounded-reasoning models, behavior is often summarized by distributions of ``levels''. We argue that such summaries conflate two conceptually distinct dimensions: a player's \emph{type}, capturing beliefs about what types their opponents might be, and the \emph{depth} of higher-order reasoning about rationality. Distinguishing these dimensions matters for interpreting experimental evidence and for understanding when cross-environment variation should be read as changes in beliefs versus changes in cognitive depth, but existing frameworks provide no language to do so. We develop a unified framework by ``lifting'' static complete-information games into incomplete-information versions in which players are explicitly uncertain about opponents' types. Within this framework, bounded reasoning about opponents' types is represented by transparent first-order belief restrictions, while (higher-order) reasoning depth is captured by bounds on belief in rationality. We analyze three benchmark instances: \emph{downward rationalizability}, a robust baseline, and two refinements, \emph{\(\mathsf{L}\)-rationalizability} and \emph{\(\mathsf{C}\)-rationalizability}, which provide epistemic foundations---with an important nuance---for classic level-\(k\) and Cognitive Hierarchy, respectively, and clarify what ``level-\(k\)'' behavior can and cannot reveal about underlying reasoning processes.
\end{niceabstract}
	
	\begin{center}
		\begin{minipage}{0.9\textwidth}
		\end{minipage}
	\end{center}
    \makeatletter
\let\@makefntext\orig@makefntext
\makeatother
	\epigraph
  {%
    ``Tatsache ist, da\ss{} eine Kalkulation der Wirkungen des  künftigen eigenen Verhaltens auf künftiges fremdes Verhalten und vice versa immer erfolgt, also jederzeit empirisch beobachtbar ist. Jedoch bricht die Kette der gemutma\ss{}ten ineinandergreifenden 'Reaktionen' verhältnismä\ss{}ig bald ab[.]''%
  }
  {Oskar \citet{morgenstern1935vollkommene}, \textit{Vollkommene Voraussicht und wirtschaftliches Gleichgewicht}, p. 343.\footnotetext{``The fact is that a calculation of the effects of one's own future behavior always rests on the expected future behavior of others, and vice versa. This can be observed empirically every time. However, the chain of surmised mutual 'reactions' breaks off comparatively soon[.]'' Translation from Frank H. Knight in print as \citet{morgenstern1976}.}}


\section{Introduction}

Bounded-reasoning models such as classic level-\(k\) \citep*{nagel1995unraveling, stahl1994experimental, stahl1995players} and Cognitive Hierarchy \citep*[CH,][]{camerer2004cognitive} are widely used\ to summarize experimental behavior through distributions of observed ``levels.'' But the meaning of a ``level'' label is actually opaque and ambiguous. The same observed choice pattern can arise either because a player believes she faces lower-type opponents---a \emph{belief-driven bound}---or because her capacity for higher-order reasoning about rationality is limited---a \emph{cognitive bound}. Conflating these channels makes experimental evidence hard to interpret, because the two channels have different comparative statics, they imply different patterns of variation across games and different responses to incentives, feedback, and framing.
Yet standard models of bounded reasoning provide no language to tell them apart. This paper provides this language by means of a unified framework in which the two mechanisms are explicit and distinct objects, making precise what the standard models leave implicit and enabling design and estimation strategies that can disentangle these two channels.

The difficulty is not merely one of classification or fit. Level-\(k\) models\footnote{In the literature, the model introduced by \citet{nagel1995unraveling} is often referred to simply as ``level-$k$.'' \citet{crawford2013structural} use ``level-$k$'' as an umbrella term encompassing both \citet{nagel1995unraveling}'s level-$k$ and CH models. To distinguish the specific model from the broader class, we use the term ``classic level-$k$'' to indicate the former. We also follow \citet{crawford2013structural} by calling this approach structural.} are structural---they represent reasoning as an algorithmic procedure and generate predictions about actions, but they do not provide explicit belief foundations. In particular, structural models typically do not specify what players believe about opponents' reasoning, only what they do. This is precisely why the belief-driven and cognitive-depth sources of bounded behavior described above are hard to separate within the standard framework. A natural way forward is to bring in tools from epistemic game theory (EGT), which formalizes strategic reasoning through hierarchies of beliefs and derives behavior from epistemic assumptions (such as rationality and common belief in rationality).\footnote{For surveys, see \citet{battigalli1999recent}, \citet{perea2012epistemic}, and \citet{dekel2015epistemic}. \citet*{battigalli2023game} offers a thorough and conceptually rich treatment permeated by EGT motivations, even if it is not about formal EGT.}  EGT shares the motivation of the bounded-reasoning literature---to capture how (boundedly) rational agents reason about others---yet the two traditions have developed largely in parallel. Moreover, EGT is often viewed as a framework suited only to ``full'' rationality, which has contributed to the perception that epistemic methods and level-\(k\) models are fundamentally incompatible.

We show that this perceived incompatibility is illusory by providing a unified framework that makes the two channels of bounded behavior explicit and separable. Thus, we turn the informal ingredients of level-\(k\) analysis into objects that can be analyzed within standard belief-based methods. The key is to take the uncertainty that bounded-reasoning models implicitly introduce---about both opponents' actions and their reasoning types---seriously as a modeling primitive. First, there is uncertainty about what actions co-players will choose, i.e., \emph{strategic uncertainty}. Second, there is uncertainty about what ``level-types'' co-players are---about how they reason, what they take as given, and which behavioral rules they follow---i.e., \emph{structural uncertainty}. Structural models are designed to speak to the former,\footnote{Admittedly, the naming convention in the literature that we follow here is somewhat confusing with respect to the ---also standard---interpretation of structural uncertainty.} but they typically do not provide a language that cleanly separates the two channels. In particular, they do not allow one to separate higher-order reasoning about rationality and action choice from reasoning about opponents' level-types, nor to study how these dimensions interact.

This separation is not only conceptually natural, but also methodologically useful. In experimental applications of level-\(k\) and CH, researchers typically infer ``types'' from observed choices and then compare estimated type distributions across games. Yet such inferred distributions often vary substantially across environments, which raises a basic interpretive question: what, exactly, does a classification as ``level-\(k\)'' reveal about the underlying reasoning process? Our framework makes this question precise because it separates two distinct mechanisms that are otherwise conflated. The same choice pattern can arise either from limited higher-order reasoning about rationality (a cognitive bound on the depth of belief in rationality) or from genuinely belief-driven bounded reasoning, in which opponents are believed to be of lower ``types''. Since the former is captured by the depth of higher-order reasoning about rationality, while the latter is encoded in our model by explicit first-order restrictions on beliefs about opponents' actions and types, the framework provides a clean language for disentangling these sources in the interpretation of ``level-\(k\)'' behavior.

We develop a tractable and conceptually clean framework that makes this separation explicit. We begin with (static) complete-information games, but ``lift'' them to incomplete-information versions in which each player is uncertain not only about co-players' actions, but also about their \emph{types}, which in our setting are reminiscent of \emph{levels} in the structural models.\footnote{To the best of our knowledge, this modeling approach is novel. \citet[p.~7]{crawford2013structural} note that ``'types' as they are called [have] no relation to private-information variables,'' whereas in our framework we explicitly treat types as private-information variables. If one takes bounded-reasoning models seriously, then players face this extraneous uncertainty about their co-players' levels. Nevertheless, \cite{strzalecki2014depth} provides a different model that makes reasoning about other players' level explicit too and we discuss the connections and differences in more detail later.} The gain from this representation is that it turns the informal ingredients of bounded-reasoning models into explicit, first-order belief restrictions about what each type believes regarding opponents' actions and types.
Assuming these restrictions are transparent\footnote{Informally, this means that the relevant restrictions are ``common knowledge'' among the players, although the finite-order implications emphasized here do not require passing to the limit of unbounded higher-order reasoning.} among players, rationality together with \((n-1)\)-th-order belief in rationality implies behavior captured by \(n\)-\(\Delta\)-rationalizability \citep{battigalli1999rationalizability, battigalli2003rationalization}. This epistemic foundation makes clear how higher-order reasoning operates within the model and thereby allows us to separate two conceptually distinct sources of bounded behavior: limitations in higher-order reasoning about rationality, that is, cognitive limits on the depth of belief in rationality, and genuinely belief-driven bounded reasoning, which in our framework is encoded directly via the imposed first-order restrictions on beliefs about opponents' actions and types. 


A further advantage of the framework is its tractability. The restrictions on beliefs and types are clean and easy to interpret, yet precise enough to deliver sharp characterizations without requiring heavy technical machinery. We view this absence of heavy technical machinery as a substantive virtue of our approach rather than a limitation. The framework is rich enough to recover and distinguish central bounded-reasoning benchmarks while remaining analytically explicit and manageable. We illustrate this flexibility and tractability through three benchmark specifications of the belief restrictions. We study ($i$) a robust baseline, \emph{downward rationalizability}, together with two refinements, ($ii$) a specification that delivers a foundation for the classic level-\(k\) model, and ($iii$) a specification that provides a counterpart for the Cognitive Hierarchy model, where some of the implications turn out to be quite different. In each case, the model is pinned down by easy-to-state and intuitive first-order restrictions on beliefs about opponents' types and, through the anchor, about the behavior of a level-\(0\) type. Throughout, we keep separate the order of higher-order reasoning about rationality, denoted by \(n\), from the player's type index, denoted by \(k\). This separation allows the analysis to distinguish limited depth of belief in rationality from belief-driven responses to perceived opponent types, and it clarifies what ``level-\(k\)'' behavior does---and does not---identify in experimental applications.\footnote{See, for example, \citet*{kneeland2015identifying}, \citet*{jin2021does}, and \cite*{chen2025measuring}.}

Going into more details, across all three instances, we impose a belief restriction we refer to as (K1), which enforces the anchor: if a type believes that her co-player might be of type 0, then conditional on that event she expects the co-player to play according to the anchor. The second main assumption concerns which co-player types a given player deems possible. While the specifics of this assumption differ across the three instances, they share the feature that each player is certain of being of a strictly higher type than her co-players. Importantly, however, our general framework is not tied to this restriction, because it also allows one to impose transparent belief restrictions under which a type \(k\) deems it possible that the co-player is of the same type, or even of a higher type. Such ``non-downward'' specifications are difficult to accommodate in standard structural formulations, but can be expressed seamlessly in our framework.\footnote{We do not pursue them here in order to keep the analysis focused, but we view them as a promising direction for future work, especially in light of recent experimental evidence suggesting that subjects may not systematically view themselves as more sophisticated than their opponents. \citep*{halevy2025magic}}  

The three instances we consider then specialize the (downward) assumption as follows.
\begin{itemize}
    \item \(\mathsf{D}\)(ownward) rationalizability: each type believes that its co-player is of a strictly lower level, but can form any belief about how these lower types are distributed.
    \item \(\mathsf{L}\)(evel-\(k\))-rationalizability: each type believes that it faces a co-player of exactly one level lower.
    \item Cognitive-Hierarchy rationalizability (\(\mathsf{C}\)-rationalizability): each type believes that co-player types are distributed according to a truncated version of a common level distribution, as in the CH model.
\end{itemize}
We then impose that these belief restrictions hold, players are rational, and that there is ($n-1$)-order belief in both rationality and these belief restrictions. As mentioned before, the behavioral implications of these epistemic assumptions yield versions of the classic and general solution concept of \(\Delta\)-rationalizability \citep{battigalli1999rationalizability, battigalli2003rationalization}, which serves as the unifying solution concept for all three instances.

In the first case, \emph{downward rationalizability} analyzed in \autoref{sec:down_rat}, our approach introduces a robust solution concept inspired by informational robustness \citep{bergemann2013robust, bergemann2016bayes, bergemann2017belief} and the Wilson doctrine in mechanism design \citep{wilson1987, bergemann2009robust}.\footnote{In a similar vein, we outline and discuss in more detail how our approach can be viewed as a robust version of the one proposed by \citet*{strzalecki2014depth} in \autoref{subsec:rel_literature} and towards the end of \autoref{sec:down_rat}, respectively.} Under the minimal belief restrictions described above, we obtain a transparent characterization in \autoref{theorem:characterization_drat}: if we require robustness across all anchors, only undominated actions survive.\footnote{Here and later, our notion of robustness---mathematically formalized via a union---is the same as in the literature on informational robustness and implementation in robust mechanism design.} This result provides a benchmark for the limits of behavioral prediction under minimal assumptions, and it connects robustness principles from mechanism design to bounded reasoning in games.\footnote{Although we do not directly address questions in mechanism design as analyzed in \cite*{de2019level}, \citet{crawford2021efficient}, \citet{kneeland2022mechanism}, or \citet*{de2023continuous}, our results in this paper---and the framework itself---can be easily extended for similar questions. We leave this avenue for future research.}

The second case, \emph{\(\mathsf{L}\)-rationalizability}, analyzed in \autoref{sec:L-rat}, provides the paper's central bridge to the classic level-\(k\) model. Specifically, \autoref{theorem:coincidence-classic} shows that up to step \(k-1\), the induced behavior is characterized by \(n\)-rationalizability in the complete-information game, whereas from step \(k\) onward it coincides exactly with classic level-\(k\) behavior. This yields a precise epistemic foundation for classic level-\(k\) reasoning within our payoff-uncertainty construction and clarifies what level-\(k\) behavior does, and does not, identify. We further show in \autoref{subsec:Lk_robust} that robustness across all anchors has bite: even with two orders of reasoning, robust level-\(k\) rationalizability can strictly refine the predictions obtained from two rounds of iterative elimination of dominated actions. Additional connections are deferred to the appendix.\footnote{The appendix shows that, after conditioning on types whose behavior is consistent with \(n\) rounds of reasoning in the complete-information game, \(\mathsf{L}\)-rationalizability recovers exactly the set of \(n\)-rationalizable actions. It also contains the comparison with \citet*{brandenburger2020two}, emphasizing that although the two constructions are conceptually distinct, both identify a sense in which level-\(k\) behavior is compatible with standard rationalizability once attention is restricted to suitable types.}

In the third case, \emph{\(\mathsf{C}\)-rationalizability} analyzed in \autoref{sec:CH-rat}, \autoref{theorem:coincidence-CH} provides a corresponding foundation for the Cognitive Hierarchy model. As in the level-\(k\) case, sufficiently high reasoning depth reproduces the classic CH predictions. The examples and theorem also reveal a key difference, however. When reasoning about rationality is limited, \(\mathsf{C}\)-rationalizability does not recover the corresponding order of complete-information rationalizability, but instead yields genuinely different implications. This already distinguishes the CH case from classic level-\(k\), and it also foreshadows the weaker consistency connection developed in the appendix. It further suggests that robustness should deliver relatively weak predictions in this case. We confirm this in \autoref{subsec:CH_robust}, where we consider a richer notion of robustness by taking the union not only across anchors but also across level distributions. Under this broader criterion, robust \(\mathsf{C}\)-rationalizability coincides with the set of undominated actions in generic games.

Before turning to the analysis, we briefly discuss the related literature and situate our contribution within it. The remainder of the paper is then organized as follows. \autoref{sec:complete_info} introduces the notation, the complete-information game as the primitive object, and the associated solution concepts. \autoref{sec:payoff_uncertainty} formalizes the ``lift'' to incomplete information. The analysis that follows is organized around three comparative benchmarks. Section~\ref{sec:down_rat} studies downward rationalizability as the robustness benchmark of the framework. Section~\ref{sec:L-rat} then provides the paper's main positive bridge to classic level-\(k\). Section~\ref{sec:CH-rat} turns to CH and shows that the same architecture has a fundamentally different implication there. Taken together, these three cases identify which features of bounded-reasoning models are robust consequences of transparent belief restrictions, which admit an exact epistemic foundation, and where the limits of that foundation lie.


\subsection{Related Literature and Contributions}\label{subsec:rel_literature}

Our paper contributes to two strands of literature that have developed largely in parallel. 
The first studies \emph{bounded strategic reasoning} through structural models such as the classic level-\(k\) and the Cognitive Hierarchy (CH) model.\footnote{\citet*{crawford2013structural} provide a comprehensive survey.} These models have been successfully applied in auctions and communication games \citep{crawford2007level, crawford2003lying, wang2010pinocchio}, and more recently in mechanism and market design \citep*{crawford2009behaviorally, crawford2021efficient, kneeland2022mechanism}. A smaller subset of this literature explicitly notes that ``bounded reasoning'' can reflect either limits on the depth of iterative reasoning or restrictions on beliefs about others' reasoning types \citep{stahl1993evolution, stahl1994experimental, ohtsubo2006depth, crawford2013structural}. However, this distinction is typically discussed informally rather than represented as separate primitives within a unified formal framework.  Our approach provides a simple formal language that clearly separates the two factors.

This differentiation has direct implications for interpreting cross-game variation in estimated types and, in particular, for the debate on whether strategic sophistication persists across games and over time \citep*{georganas2015persistence, agranov2012beliefs, chen2025measuring}. The key point is an identification one: without auxiliary information that disciplines beliefs about opponents, observed choices generally do not reveal whether ``level-\(k\)'' behavior reflects a belief-driven boundor a cognitive bound. As a result, shifts in estimated type distributions across games or treatments could reflect changes in cognitive sophistication and changes in first-order beliefs induced by the environment (information conditions, framing, or the perceived subject pool).\footnote{This perspective also helps organize additional mechanisms emphasized in recent work, such as ambiguity-based explanations \citep*{cooper2024consistent}, as operating through the belief channel or interacting with it.}


Our framework therefore points toward empirical strategies that separate the two channels rather than treating estimated ``levels'' as a direct measure of sophistication. Designs that vary information about the opponent pool or provide feedback about others' past play primarily shift subjects' first-order beliefs about opponents' types and actions, whereas time pressure, cognitive load, training, and strong incentives more naturally affect the depth of higher-order reasoning about rationality.
Belief-control designs are especially clean in this regard. For example, \citet{chen2025measuring} pair subjects with opponents that are commonly known to be fully rational, thereby standardizing expectations about others and making residual deviations more naturally interpretable as limits in higher-order reasoning. For estimation and reporting, our framework provides a principled basis for ($i$) incorporating belief measurements or belief-disciplining features where feasible, ($ii$) checking robustness to anchor assumptions that encode first-order beliefs, and ($iii$) organizing cross-environment comparisons around belief shifts versus depth shifts even when point identification is unavailable.\footnote{These implications also connect to a broader literature on the cognitive and motivational underpinnings of reasoning depth \citep*{alaoui2016endogenous, alaoui2020reasoning, halevy2025magic} and to empirical investigations of when and why level-$k$ behavior arises \citep{jin2021does}.}

The second strand is the literature on \emph{epistemic game theory (EGT)}. EGT formalizes strategic reasoning through explicit hierarchies of beliefs, studying how rationality and epistemic assumptions such as common belief in rationality shape predictions, with rationalizability as the canonical behavioral counterpart \citep{brandenburger1987rationalizability, tan1988bayesian, battigalli1999hierarchies, friedenberg2021iterated}. A related branch of work uses epistemic tools to model \emph{genuine} bounds on reasoning by restricting the epistemic depth agents can attain or the procedures they can perform \citep*{sakovics2001games, heifetz2018robust, kets2021bounded, heifetz2019robust, germano2020uncertain}. Our paper connects these two branches. We do not impose an ex ante bound on higher-order reasoning within the model; instead, bounded-reasoning behavior arises \emph{as-if} from transparent first-order belief restrictions, while the order of higher-order belief in rationality is tracked explicitly by the parameter \(n\). If desired, genuine depth bounds can be incorporated in our framework by imposing an upper bound on \(n\), though our baseline results do not assume limited epistemic depth.

Closest to our approach in spirit are \citet{strzalecki2014depth} and \citet*{brandenburger2020two}. The former makes players' uncertainty about others' reasoning explicit by adopting \citeauthor{harsanyi1967games}'s (\citeyear{harsanyi1967games}) framework of incomplete information. The latter examine to what extent classic level-\(k\) analysis sheds light on bounded reasoning in games and, in doing so, address the open question of whether standard complete-information epistemic models can rationalize level-\(k\) behavior. We pursue a complementary question and take a different route by ``lifting'' complete-information games to incomplete-information settings. In this way, we obtain a transparent mapping between structural primitives (levels/types and anchors) and epistemic restrictions, and can study robustness across anchors and reasoning orders within a unified framework. The difference between our framework and that of \citet{strzalecki2014depth} parallels the difference between games with payoff uncertainty and Bayesian games. Our approach also connects to robustness considerations in mechanism design \citep{wilson1987, bergemann2005robust, bergemann2009robust, bergemann2013robust, bergemann2016bayes, bergemann2017belief}, and to related treatments of belief restrictions in \citet*{artemov2013robust} and \citet{ollar2017full, ollar2022efficient, ollar2023network}. Unlike the local robustness analyses of \citet{weinstein2007structure} or \citet{murayama2020robust}, our focus is on informational robustness across all anchors and reasoning levels. Finally, \citet{schipper2024level} use ideas inspired by EGT to study versions of level-\(k\) models for sequential games. Our analysis, although focused on static games, is not only inspired by EGT but also crucially relies on the formal epistemic results in \citet{battigalli2003rationalization}.

    \section{The Primitive Object: A Complete Information Game}\label{sec:complete_info}

\subsection{Notation}
We use \(\mathbb{N}_0\) to denote the set of natural numbers and \(\mathbb{N}\) the set of positive integers, that is, \(\mathbb{N}_0 = \{0,1,...\}\) and \(\mathbb{N} = \mathbb{N}_0 \setminus \{0\}\). Given a set \(X\), we use \(\Delta(X)\) to denote the set of all probability measures on \(X\) (with respect to a \(\sigma\)-algebra that is obvious in the context), and \(\Delta_+(X)\) the set of such measures with full support.

\subsection{The Game}
 Without essential loss of generality, we focus on \(2\)-player games,\footnote{We say this is without essential loss, because with more than two players, all our analysis carries through. However, the usual issues regarding correlated conjectures---and potentially correlated anchors---have to be considered.} which are commonly used in experimental research.\footnote{Many seminal experiments on classic level-\(k\) reasoning adopt an \(n\)-player setting, for example, \cite*{nagel1995unraveling, ho1998iterated, bosch2002one}, because they adopted \citeauthor{keynes1936}'s (\citeyear{keynes1936}) beauty contest game. However, as noted by \citet{crawford2013structural}, such general designs make it difficult to isolate the influence of one player's behavior on others. For this reason, much of the subsequent literature has focused on \(2\)-player games, an assumption we follow here.} 
 Accordingly, we consider a finite static game \( G = \langle I, (A_i, \pi_i)_{i \in I} \rangle \), where \( I = \{1,2\} \), and for each \( i \in I \), \( A_i \) represents the finite set of actions available to player \( i \), and \( \pi_i : A_1 \times A_2 \to \mathbb{R} \) is \(i\)'s payoff function. Thus, we implicitly analyze a \textit{reduced} game directly. This assumption is, for our purposes, also without loss of generality.\footnote{For the formal definition and the distinction between an (unreduced) game and a reduced game, see \citet*[Chapter 2]{battigalli2023game}. However, we want to stress that for concrete applications, particularly in experimental settings, it may be beneficial to begin with an unreduced game.}

    In this complete information setting, the primitive uncertainty is about the opponent's choice; hence a \emph{conjecture} of player \(i\) is given by \(\nu^i \in \Delta(A_{-i})\).\footnote{Note that the index \(i\) for conjecture \(\nu^i\) appears in the superscript rather than the subscript. This notation distinguishes between two fundamentally different concepts: superscripts denote variables related to \emph{other agents'} behavior and, as in games with payoff uncertainty, types, which are uncertain from any given player's perspective, while subscripts, as in \(\pi_i\) and \(a_i\), indicate either the player's own features or variables under their direct control. For the anchor \(p_i\) the index is in the subscript because, first and foremost, \(p_i\) describes the \emph{behavior} of a ``non-sophisticated'' player.} To distinguish these conjectures from those in a derived incomplete information game, we will refer to them as \emph{\(G\)-conjectures}. Given a  \(G\)-conjectures \(\nu^i\), an expected-payoff maximizer chooses an action that maximizes her expected payoff under \(\nu^i\), which is captured by the \emph{best-reply correspondence} \( r_i : \Delta(A_{-i}) \rightrightarrows A_i \), where  
\(
r_i(\nu^i) := \argmax_{a_i \in A_i}
\sum_{ a_{-i} \in A_{-i}} \nu^i( a_{-i}) \pi_i(a_i, a_{-i}).
\)
An action \(a_i\) is called \emph{justifiable} if and only if \(a_i \in r_i(\nu^i)\) for some conjecture \(\nu^i \in \Delta(A_{-i})\). In this case, we say \(\nu^i\) \emph{justifies} \(a_i\).

For complete information games, we give the definitions of rationalizability and two well-known solution concepts that are intended to capture level-\(k\) reasoning in the literature.

\subsection{Rationalizability}\label{subsec:rat}
Let \(\left( \rat^n := \prod_{i \in I}\rat_i^n \right)_{n \in \mathbb{N}_0}\) denote the sequence of action-profile sets obtained through iterated elimination of (strictly) dominated actions. That is, for each \(i \in I\), \(\rat_i^0 = A_i\), and for each \(n \geq 0\), \[
R^{n+1}_i :=
\Bigg\{a_i \in R_i^n
\Bigg\vert
\nexists \,\alpha_i \in \Delta(R_i^{n}) , \, \forall a_{-i} \in R_{-i}^n, \, \sum_{a_i' \in A_i} \alpha_i(a_i)\pi_i(a'_i,a_{-i}) > \pi_i(a_i,a_{-i})\Bigg\}.
\]
An action \(a_i\) is \emph{iteratively undominated} if and only if \(a_i \in R_i^\infty := \cap_{n \in \mathbb{N}}R_i^n\). 
It is well known that for finite games, \(R_i^1\) coincides with the set of justifiable actions and that iterated elimination of dominated actions coincides with \emph{rationalizability}. Thus, we let  
\(
\rat^\infty := \prod_{i \in I} \rat_i^\infty
\)
as the set of \emph{rationalizable} action profiles, which correspond to the behavioral implications of rationality and common belief in rationality. For any \(n \geq 0\), we refer to actions in \(R_i^n\) as the \(n\)-rationalizable actions of player \(i\) and we will use similar naming convention for other versions of rationalizability defined below.

\subsection[Classic Level-k]{Classic Level-\(k\)}\label{subsec:static_level-k}
The classic level-\(k\) model, originating with \citet{nagel1995unraveling}, assumes that each level-\(k\) player believes their opponent to be of level-\((k-1)\), and that a level-\(0\) player's behavior is determined by an exogenous anchor. An \emph{anchor} is given by \( p = (p_1, p_2) \in \Delta(A_1) \times \Delta(A_2) \) and we assume---somewhat implicitly---stochastic independence across the players for their anchors throughout.\footnote{In principle, one could allow for a correlated anchor. However, we follow the standard practice in the literature---e.g., \citet*{brandenburger2020two}---and assume independence. While this assumption does not affect the two-player case, it may become significant in games with more than two players.} Fixing an anchor \(p\), the classic level-\(k\) solution \(\left(\lk^k := \prod_{i \in I} \lk[i]^k\right)_{k \in \mathbb{N}}\) is inductively defined as follows.\footnote{The behavioral economics literature often assumes a specific tie-breaking rule when there are multiple (optimal) actions for a given level-\(k\). Here, following \citet{brandenburger2020two}, we do not impose such an assumption.}

\begin{enumerate}[leftmargin=!, labelwidth=!, labelsep=0.5em, align=left]
    \item[\textbf{Step \(1\)}.] For every \(i \in I\), \(\lk[i]^1 := r_i(p_{-i})\).

    \item[\textbf{Step \(k+1\)}.] For every \(i \in I\) and \(a_i \in A_i\), \(a_i \in \lk[i]^{k+1}\) if and only if there exists \(\nu^i \in \Delta(A_{-i})\) such that:
    \begin{itemize}[nosep]
        \item[(i)] \(a_i \in r_i(\nu^i)\), and
        \item[(ii)] \(\nu^i\bigl(\lk[-i]^k\bigr) = 1\).
    \end{itemize}
\end{enumerate}
We refer to \(\lk[i]^{k}\) as the \emph{level-\(k\) behavior of player \(i\)}, and \(\lk^{k}\) as the \emph{level-\(k\) behavior}. The sequence \((\lk^{k})_{k \in \mathbb{N}}\) is called the \emph{classic level-\(k\) solution}.

\subsection{Cognitive Hierarchy Model}\label{subsec:static_CH}

An alternative approach to capture level-\(k\) reasoning is the cognitive hierarchy (CH) model introduced by \citet*{camerer2004cognitive}. The CH model shares the spirit of level-\(k\) reasoning with the classic level-\(k\) model by assuming that each player assumes that they are more sophisticated than their co-player and a level-\(0\) player's behavior is determined by an exogenous anchor. The difference is that, in addition, CH assumes that there is a common distribution \(f\) over strategic sophistications (levels), and each player, given their own level \(k\), believes the co-player's sophistication is random and distributed according to the normalization of the truncated \(f\) up to \(k-1\). Formally, given an anchor \( p = (p_1, p_2) \in \Delta(A_1) \times \Delta(A_2) \) and a \textit{level distribution} \(f \in \Delta_+(\mathbb{N}_0)\), for every \(k\in \mathbb{N}\), we define \(f^k(t) :=  \frac{f(t)}{\sum_{t'=0}^{k-1}f(t')}\) for every \(t = 0, \ldots ,k-1\).
The \emph{CH solution} \((CH^k[p,f]:=\prod_{i \in I}CH^k_i[p,f])_{k \in \mathbb{N}}\) is defined inductively as follows:
		\begin{enumerate}
			\item[] \textbf{Step \(1\).} For every \(i \in I\), \(CH^1_i[p,f] := r_i(p_{-i})\).
			
			\item[] \textbf{Step \(k+1\)}. For every \(i \in I\) and \(a_i \in A_i\), \(a_i \in CH^{k+1}_i[p,f]\) if and only if there exists \(\nu^i \in \Delta(\mathbb{N}_0 \times A_{-i} )\) such that:
			\begin{itemize}[nosep]
				\item[(i)] \(a_i \in r_i(\marg_{A_{-i}}\nu^i)\),
                \item[(ii)] \(\marg_{\mathbb{N}_0}\nu^i = f^k\), 
                \item[(iii)] \(\nu^i\left(\left. \cdot \right\vert 0 \right) = p_{-i}\), and 
				\item[(iv)] for \(0 < t \leq k\), \(
                \nu^i\big(CH^t_{-i}[p,f] \big\vert t \big) = 1\)
			\end{itemize}
		\end{enumerate}

We refer to \(CH_i^{k}[p,f]\) as the \emph{CH level-\(k\) behavior of player \(i\)} and \(CH^{k}[p,f]\) as the \emph{CH level-\(k\) behavior}. The sequence \((CH^{k}[p,f])_{k \in \mathbb{N}}\) is called the \emph{CH level-\(k\) solution}.


\section{The Derived Game With Payoff Uncertainty}\label{sec:payoff_uncertainty}
In a complete information game, a player's uncertainty concerns the co-player's action. However, in the two level-\(k\) models introduced---and more saliently in the CH model---there is additional uncertainty regarding the co-player's level of reasoning.\footnote{Technically, there is also uncertainty about the reasoning process itself. We do not model this explicitly, but leave it implicit. Epistemic game theory, of course, makes this uncertainty explicit.} A standard tool for incorporating such additional uncertainty is the use of a \emph{game with payoff uncertainty}.\footnote{Equivalently, one might refer to this as a \emph{game with incomplete information}. However, we avoid this terminology, as the literature employs it to denote varying concepts.} Naturally, we want the game with payoff uncertainty to be grounded in the complete information game~\(G\).

Thus, given the complete information game \( G =\langle I, (A_i, \pi_i)_{i \in I} \rangle\), we extend the framework by introducing types associated with levels for each player and defining the derived game with payoff uncertainty, denoted by \( \hat{G} = \langle I, (A_i, \Theta_i, u_i)_{i \in I} \rangle \), as follows:
\begin{itemize}[nosep]
    \item for every \( i \in I \), let \( \Theta_i := \{\theta_{i,0}, \theta_{i,1},...\} = \{\theta_{i,k} : k \in \mathbb{N}_0\} \), where each \( \theta_{i,k} \) is referred to as the \emph{level-\(k\) type} of player \( i \), and
    
    \item for every \( i \in I \), given \(\theta_{i,k} \in \Theta_i \) and \( a \in A := A_1 \times A_2 \), the payoff function is defined as

    \begin{equation*}
				u_i(\theta_{i,k}, a) = 
				\begin{cases}
					0 & \text{if } k = 0, \\
					\pi_i(a) & \text{otherwise}.
				\end{cases}    
			\end{equation*}
            
\end{itemize}

Here, the ``authentic'' payoff uncertainty arises from the presence of a level-\(0\) type: for all other types, the utility function \( u_i \) depends solely on \( a \), whereas for the level-\(0\) type, \( u_i \) remains constant regardless of the action profile. Within this framework, we can also assume rationality for players of type \(0\). Such a player may randomize their choices not due to a lack of strategic sophistication, but because her payoff remains constant across all actions. While this assumption is not essential, it simplifies the model and ensures a unified conceptual approach.\footnote{In accordance with the literature, we do not seek to explain the anchor, i.e. the behavior of the level-\(0\) types. See \autoref{remark:0-types}.} Indeed, this modelling choice makes the derived game \(\hat{G}\) independent of any anchor.\footnote{Nevertheless, all our results would go through if we would make \(\hat{G}\) anchor specific in the sense that level-\(0\) types have payoffs that make those and only those actions optimal which are in the support of the anchor.}

Furthermore, we also wish to emphasize the importance of careful interpretation: \( \theta_{i,k} \) serves only as a label for a cognitive state; the \( k \)  does not \emph{per se} convey any inherent relationship to strategic sophistication or cognitive abilities. Thus, \( k \) should be understood purely as an ``index'' label. Later, as shown in \autoref{prop:bounded_behavior}, the behavioral implications of seemingly limited strategic sophistication of each type will be derived from the epistemic condition to be stated.


As foreshadowed before, in the game with payoff uncertainty \( \hat{G} \), the primitive uncertainty is about the type of the co-player as well as the choice; hence, a \textit{conjecture} of player \( i \) is defined as a distribution over their co-player's type-action combinations, i.e., \( \mu^i \in \Delta(\Theta_{-i} \times A_{-i}) \). To differentiate these conjectures, from \(G\)-conjectures defined earlier for the complete information game, we will refer to these ones as \(\hat{G}\)-conjectures. Furthermore, to ease notation burden, with a slight abuse of notation, for each \(\hat{G}\)-conjecture \( \mu^i \) and each \( \theta_{-i,k} \in \Theta_{-i} \), we use \( \mu^i(\theta_{-i,k}) \) to denote the probability of \( \theta_{-i,k} \) with respect to the marginal distribution of \( \mu^i \) on \( \Theta_{-i} \): formally,  
\(
\mu^i(\theta_{-i,k}) = \sum_{a_{-i} \in A_{-i}} \mu^i(\theta_{-i,k}, a_{-i}).
\) 
For any given \(\theta_{-i,k}\) such that \( \mu^i(\theta_{-i,k}) > 0 \), we denote the distribution induced by \( \mu^i \) on \( A_{-i} \), a \(G\)-conjecture conditional on \( \theta_{-i,k} \), by \( \mu^i(\cdot \mid \theta_{-i,k}) \). That is, for each \( a_{-i} \in A_{-i} \),  
\(
\mu^i(a_{-i} \mid \theta_{-i,k}) = \frac{\mu^i(\theta_{-i,k}, a_{-i})}{\mu^i(\theta_{-i,k})}.
\)

Following \citet{battigalli1999rationalizability} and \citet{battigalli2003rationalization}, we consider a class of solution concepts for the game with payoff uncertainty, parametrized by (first-order) \emph{belief restrictions} \(\Delta = (\Delta^1, \Delta^2)\), where \(\Delta^i = \left(\Delta^{\theta_{i,k}}\right)_{k \in \mathbb{N}}\) for each \(i \in I\), and each \(\Delta^{\theta_{i,k}} \subseteq \Delta(\Theta_{-i} \times A_{-i})\) is non-empty and specifies the belief restrictions for player \(i \in I\) under type \(\theta_{i,k}\). Fixing belief restrictions \(\Delta\), we define \emph{\(\Delta\)-rationalizability} through a sequence \(\left(\mathsf{R}_\Delta^n:= \prod_{i \in I} \mathsf{R}_{i,\Delta}^n\right)_{n \in \mathbb{N}_0}\) recursively as follows.
\begin{enumerate}[leftmargin=!, labelwidth=!, labelsep=0.5em, align=left]
    \item[\textbf{Step \(0\).}]
    For every \(i \in I\), set \(\mathsf{R}_{i,\Delta}^{0} := \Theta_{i} \times A_{i}\).
    
    \item[\textbf{Step \(n+1\).}]
    For every \(i \in I\), \(k \geq 1\), and \((\theta_{i,k}, a_{i}) \in \mathsf{R}_{i,\Delta}^{n}\), we have \((\theta_{i,k}, a_{i}) \in \mathsf{R}_{i,\Delta}^{n+1}\) if and only if there exists \(\mu^{\theta_{i,k}} \in \Delta^{\theta_{i,k}}\) such that:
    \begin{enumerate}[nosep]
        \item \(a_{i} \in r_i(\marg_{A_{-i}}\mu^{\theta_{i,k}})\), and
        \item \(\mu^{\theta_{i,k}}(\rat_{-i,\Delta}^{n}) = 1\). 
    \end{enumerate}
\end{enumerate}
Finally, for every \(i \in I\), we define  
\(
\mathsf{R}_{i,\Delta}^{\infty} :=\bigcap_{n\geq 0} \mathsf{R}_{i,\Delta}^{n}
\)
as the set of \emph{\(\Delta\)-rationalizable} type-action pairs for player \(i\). The set of profiles of type-actions pairs that are \(\Delta\)-rationalizable is denoted by  
\(
\mathsf{R}_\Delta^{\infty} := \mathsf{R}_{1,\Delta }^{\infty} \times  \mathsf{R}_{2,\Delta}^{\infty}.
\)
Furthermore, for every \( n \in \mathbb{N}_0 \cup \{\infty\} \) and \( \theta_{i,k} \in \Theta_i \), we define \( \mathsf{R}^n_{i,\Delta}(\theta_{i,k}) := \{a_i \in A_i : (\theta_{i,k}, a_i) \in  \mathsf{R}^n_{i,\Delta} \} \), i.e., the section of \( \mathsf{R}^n_{i,\Delta} \) at \( \theta_{i,k} \). Thus, this defines a correspondence \(  \mathsf{R}^n_{i,\Delta}: \Theta_i \rightrightarrows A_i\) for every player \(i \in I\) and \(n \in \mathbb{N}_0 \cup \{\infty\}\). Henceforth, we will use an equivalent notation for any subset of \(\Theta_i \times A_i\).

The motivation behind the procedure is as follows. First, in epistemic game theory, two fundamental assumptions are \emph{rationality} and \emph{common belief in rationality}. Rationality requires that a player maximizes their expected payoff given their conjecture. \emph{Common belief} in an event in an event means that everyone believes the event, everyone believes that everyone believes event, and so on, ad infinitum. Beyond these two fundamental assumptions, we are also interested in situations where the restrictions on first-order beliefs imposed by \(\Delta\) are not only in place but are also \textit{transparent} to all players.\footnote{Again, we are informal here, but in our static setting, transparency can be equated with common belief in these belief restrictions. \citet[Section 9.b]{battigalli2012forward} provides further details.} \citet{battigalli1999rationalizability} and \citet{battigalli2003rationalization} characterize the behavioral implications of rationality, common belief in rationality, and transparency of arbitrary (first-order) belief restrictions through the \(\Delta\)-rationalizability procedure defined above.\footnote{Here, our focus is on characterizing the behavioral implications of epistemic conditions. Therefore, we provide only an informal and intuitive description of events satisfying these epistemic conditions. \autoref{subsec:rel_literature} provides references for more formal treatments.} 

\begin{remark}[Behavior of \(0\)-types]\label{remark:0-types}
    Note that our definition of \(\Delta\)-rationalizability does not impose any restrictions on the predicted play of \(0\)-types. Although this is somewhat implicit in the given definition, it is consistent with the underlying game with payoff uncertainty, because \(0\)-types have a constant payoff function.
\end{remark}

By imposing belief restrictions that capture the spirit of level-\(k\) models, our framework allows us to cleanly separate two interpretations of level-\(k\) reasoning commonly conflated in the literature: depth of reasoning and restrictions on beliefs. The former corresponds to the number of iterations \(n\) in \(\Delta\)-rationalizability, while the latter is encoded in each type \(\theta_{i,k}\). This clear separation enables us to identify the causal drivers of strategic choices: does a decision reflect genuinely bounded reasoning---where the inference process terminates at iteration \(n\)---or is it driven by beliefs about the opponent's level? Disentangling these factors has significant implications for both empirical and theoretical research. Experimentalists can more accurately attribute observed behavior to its source, and theorists can develop models and design mechanisms that account for both reasoning depth and belief restrictions. Many of our subsequent results hinge on this crucial conceptual distinction.




\section{Downward Rationalizability}\label{sec:down_rat}
In this section, we characterize behavior consistent with level-\(k\) reasoning through specifying restrictions \(\Delta\) which captures precisely the logic of level-\(k\) reasoning and studying the corresponding \(\Delta\)-rationalizability.
Fix an anchor \( p \in \Delta(A_1) \times \Delta(A_2) \). 
For every \( i \in I \) and \( k \in \mathbb{N} \), let \( \Delta^{\theta_{i,k}} \subseteq \Delta(\Theta_{-i} \times A_{-i}) \) denote the set of \(\hat{G}\)-conjectures satisfying the following conditions:  
(K1) Type-0 co-players' actions are distributed according to the anchor, and  
(K2) only co-players' levels of strictly lower order are deemed possible.  Note that (K1) imposes a belief restriction on endogenous objects only, whereas (K2) is a restriction
on exogenous beliefs only.\footnote{See Sections 1.2 and 8.3 in \citet{battigalli2023game} for the distinction between \emph{endogenous} and \emph{exogenous} variables in game theory. In particular, (K1) restricts beliefs about the (endogenous) behavior of type-\(0\) players using the (exogenous) anchor \(p\). In contrast, (K2) restricts beliefs about which co-player (exogenous) types are considered possible.}   
Formally, \( \mu^{\theta_{i,k}} \in \Delta^{\theta_{i,k}} \) if and only if
\begin{enumerate}[nosep]
    \item[\textup{(K1).}] \( \mu^{\theta_{i,k}}(\theta_{-i,0}) > 0 \implies \mu^{\theta_{i,k}}(\cdot \mid \theta_{-i,0}) = p_{-i} \), and
    \item[\textup{(K2).}] \( \supp \marg_{\Theta_{-i}} \mu^{\theta_{i,k}} \subseteq \{\theta_{-i,0}, \theta_{-i,1}, \ldots, \theta_{-i,k-1}\}. \)
\end{enumerate}
The resulting belief restrictions yield a version of \(\Delta\)-rationalizability, which we call \emph{downward rationalizability}, and the corresponding downward-rationalizable actions of player \(i\) will be denoted by \( \mathsf{D}_{i,p}^\infty \). Equivalent notation will be used for other related objects.

Intuitively, downward rationalizability can be viewed as a general and robust formalization of the intuitive ideas underlying bounded‐reasoning models, including the classic level-\(k\) model and the CH model. It is more general than the classic level-\(k\) model, but shares with the CH model the feature that a player of type \(k\) is not restricted to considering only co-players of type \(k-1\) as possible; in principle, any lower-order type could be deemed possible. Furthermore, downward rationalizability can be viewed as a robust version of the CH model, as it does not rely on (nor require) the specification of a common level distribution \(f \in \Delta_+(\mathbb{N}_0)\).
In this vein, downward rationalizability is robust in a sense similar to the notion employed in robust mechanism design \citep{bergemann2009robust, bergemann2017belief}.\footnote{Nevertheless, downward rationalizability is not entirely belief-free, as it retains minimal belief restrictions to capture the basic intuition of level-\(k\) models. See \autoref{subsec:rel_literature} for references to robust mechansim design with belief restrictions, which are not enirely belief-free.}

Note that no restrictions are imposed on the conjectures of a \(0\)-type. Since the payoff function of a \(0\)-type is constant, their optimal behavior is independent of their conjecture.\footnote{See \autoref{remark:0-types}.} Additionally, for \(1\)-types, there exists a \emph{unique} \(\hat{G}\)-conjecture that satisfies the belief restrictions. Specifically, (K2) ensures that any \(\hat{G}\)-conjecture for a \(1\)-type assigns positive probability only to \(0\)-types of their co-player, which---in other words---says that the marginal on \(\Theta_{-i}\) is unique. Overall uniqueness of the \(\hat{G}\)-conjecture follows then directly from (K1). However, for a player of type \( k > 1 \), these conditions impose no restrictions on their conjecture about a co-player with non-zero lower types.

First, we establish that the behavior of a player with type \(\theta_{i,k}\) is \emph{as if} she reasons at most \(k\) steps. So far, \(\theta_{i,k}\) has been merely a label for a type. However, with the following result, we establish that the number \(k\) is ``behaviorally'' equivalent to an upper bound on strategic sophistication: a player with type \(\theta_{i,k}\) behaves \emph{as if} she reasons at most \(k\) steps because it is unnecessary to reason beyond this depth. This upper bound on reasoning depth follows from the three epistemic assumptions, especially the transparency of (K1) and (K2). Note that in our model, all types are capable of reasoning unboundedly. However, they do not need to, due to the restriction on their first-order beliefs. Thus, in our model, players behave as if they stop reasoning---not because of a cognitive limitation, but because they believe that their opponent will behave as if they reason only up to lower levels. Thus, our model is in line with the interpretation of \cite{ohtsubo2006depth} and can be seen as a formalization of the view of \citet[Footnote 17]{crawford2013structural}: ``In our view, people stop at low levels mainly because they believe others will not go higher, not due to cognitive limitations[.]''

\begin{proposition}[Bounded Reasoning]\label{prop:bounded_behavior}
	For every \(i \in I\), every \( k \in \mathbb{N}_0 \), and every \( n \geq k \),  
	\(
	\mathsf{D}^n_{i, p}(\theta_{i,k}) = \mathsf{D}^k_{i, p}(\theta_{i,k}).
	\)
\end{proposition}

\begin{proof}
	We prove this proposition by (strong) induction on \( k \). The base step for \( k = 0 \) follows from the observation that player \( i \)'s payoff for this type is constant, and therefore \( \mathsf{D}^n_{i, p}(\theta_{i,0}) = \mathsf{D}^0_{i, p}(\theta_{i,0}) = A_i \) for every \( n \geq 0 \).
	
	Now, assume that the statement is true for some \( k > 0 \), all lower integers, and all players. We want to prove that the statement holds for all \( n \geq k \), which we prove by strong induction too, but on \(n\). The base case here, with \( n = k \), is trivial. Thus, we prove that the statement holds for \( n+1 \) with \( n > k \), assuming it holds true for all lower integers and all players.
	
	First, note that the \( \subseteq \)-inclusion always holds by definition. For the other direction, fix \( a_i \in \mathsf{D}^k_{i, p}(\theta_{i,k}) = \mathsf{D}^n_{i, p}(\theta_{i,k}) \), where the equality follows from the strong induction hypothesis (on \(n\)). Therefore, there exists a justifying \(\hat{G}\)-conjecture \( \mu^{\theta_{i,k}} \in \Delta^{\theta_{i,k}} \) such that \( \mu^{\theta_{i,k}}(\mathsf{D}_{-i,p}^{n-1}) = 1 \). 
	
	By the strong induction hypothesis, \( \mathsf{D}_{-i,p}^{n-1}(\theta_{-i,t}) = \mathsf{D}_{-i,p}^{k}(\theta_{-i,t}) = \mathsf{D}_{-i,p}^{n}(\theta_{-i,t}) \) for all \( t < k \). Combining this observation with \( \mu^{\theta_{i,k}} \in \Delta^{\theta_{i,k}} \), we obtain \( \mu^{\theta_{i,k}}(\mathsf{D}_{-i,p}^{n}) = 1 \). Thus, we conclude that \( a_i \in \mathsf{D}^{n+1}_{i, p}(\theta_{i,k}) \).
\end{proof}

Furthermore, under these restrictions, we obtain a monotonicity result for all levels \(k\geq 1\): the higher the level \( k \), the larger---by set inclusion---the set of downward rationalizable actions associated with type \(\theta_{i,k}\), which directly follows from the observation that \(\Delta^{\theta_{i,k}} \subseteq \Delta^{\theta_{i,k+1}}\) holds for all \(\theta_{i,k}\) with \(k \geq 1\). 

\begin{proposition}[\(k\)-Monotonicity]\label{prop:monotone}
	For every \( i \in I \), every \(n \in \mathbb{N}_0 \), and every \(k \in \mathbb{N} \)
	\[
	\mathsf{D}^n_{i, p}(\theta_{i,k}) \subseteq
    \mathsf{D}^n_{i, p}(\theta_{i,k+1}),
	\]  
	and, furthermore,  
	\(
	\mathsf{D}^\infty_{i, p}(\theta_{i,k}) \subseteq
    \mathsf{D}^\infty_{i, p}(\theta_{i,k+1}).
	\)
\end{proposition}

The monotonicity result of \autoref{prop:monotone} suggests that the label \(k\) of the type \(\theta_{i,k}\) cannot be equalized with the step \(k\) in the usual rationalizability argument in the corresponding complete information game: in the latter, higher-order reasoning \emph{refines} the behavioral implications. Indeed, as pointed out before, it is crucial and a desirable feature of our model to differentiate between types indicative of level-\(k\), and the actual order of reasoning that we denote with \(n\) usually. The following example illustrates the difference.

\begin{example}\label{example: downward-rationalization}
Consider the game shown in \autoref{tab:iterated_elimination_example}. Applying iterated elimination of dominated actions to the complete information game generates a sequence \((R^n)_{n \in \mathbb{N}_0}\) such that
\(\rat^1 = A_1 \times \{l, c\}\), \(\rat^2 = \{U, M \} \times \{l, c\}\), \(\rat^3 = \{U, M \} \times \{l\}\), and \(\rat^4 = \{U \} \times \{l\} = \rat^\infty.\)
\begin{table}[ht!]
    \centering
    \small
    \caption{A game}
    \label{tab:iterated_elimination_example}
    \renewcommand{\arraystretch}{0.75} 
    \setlength{\tabcolsep}{10pt} 
    \begin{tabular}{c ccc}
        \toprule
        Player 1 \textbackslash \, 2 & \textbf{\(l\)}  & \textbf{\(c\)} & \textbf{\(r\)}  \\ 
        \midrule
        \textbf{\(U\)}  & \(3, 2\) & \(2, 1\) & \(1, 0\)  \\
        \textbf{\(M\)}  & \(2, 2\) & \(3, 1\) & \(2, 0\)  \\
        \textbf{\(D\)}  & \(1, 1\) & \(1, 2\) & \(3, 0\)  \\ 
        \bottomrule
    \end{tabular}
\end{table}

Now, consider the anchor \(p = (\delta_D, \delta_r)\), where \(\delta_x\) denotes the Dirac measure assigning probability \(1\) to action \(x\). Note that both \(D\) and \(r\) are iteratively dominated. The corresponding downward rationalizability iterations is presented in \autoref{tab:downward_rationalization}. Each column \(n\) lists the actions that survives until the \(n\)-th step in \(\Delta\)-rationalizability, namely \(n\)-downward-rationalizable actions, for each type. For example, in the column for \(n=1\), one can see that \(\mathsf{D}_{1,p}^1(\theta_{1,1}) = \{D\}\), that is, only action \(D\) is \(1\)-downward rationalizable for player 1's type \(1\); similarly, by looking at the column for \(n=2\), one can see that \(\mathsf{D}_{2,p}^2(\theta_{2,1}) = \{M,D\}\), meaning that both \(M\) and \(D\) are \(2\)-downward rationalizable for player 1's type \(2\). To see the procedure underlying downward rationalizable, it is instructive to read the table column by column.

\begin{table}[ht!]
    \centering
    \small
    \caption{Downward rationalizability for \(p = (\delta_D, \delta_r)\)}
    \label{tab:downward_rationalization}
    \renewcommand{\arraystretch}{0.9} 
    \setlength{\tabcolsep}{7pt} 
    \begin{tabular}{c ccccc}
        \toprule
        \multirow{1}{*}{\(\mathsf{D}_{1,p}^n(\theta_{1,k}), \mathsf{D}_{2,p}^n(\theta_{2,k}) \)} & \multirow{2}{*}{} & \multicolumn{4}{c}{\(\mathbf{n \rightarrow}\)} \\  
        \cmidrule(lr){3-6}
        \multirow{1}{*}{\(\mathbf{k \downarrow}\)}  & & \textbf{1}  & \textbf{2} & \textbf{3} & \textbf{\(\geq 4\)}   \\ 
        \midrule
        \textbf{1}  & & \myalign{c}{\{D\} , \{c\}} & \myalign{c}{\{D\} , \{c\}} & \myalign{c}{\{D\} , \{c\}} & \myalign{c}{\{D\} , \{c\}} \\
        \textbf{2}  & & \myalign{c}{A_1 , \{l, c\}} & \myalign{c}{\{M, D\} , \{c\}} & \myalign{c}{\{M, D\} ,\{c\}} & \myalign{c}{\{M, D\} , \{c\}}  \\
        \textbf{3}  & & \myalign{c}{A_1 , \{l, c\}} & \myalign{c}{A_1 , \{l, c\}} & \myalign{c}{\{M,D\} , \{l,c\}} & \myalign{c}{\{M,D\} , \{l,c\}} \\
        \textbf{\(\geq 4\)} & & \myalign{c}{A_1 , \{l, c\}} & \myalign{c}{A_1 , \{l, c\}} & \myalign{c}{A_1 , \{l, c\}} & \myalign{c}{A_1, \{l,c\}} \\ 
        \bottomrule
    \end{tabular}
\end{table}

Next, consider the anchor \(p^\prime = (\delta_U, \delta_l)\), which corresponds to the unique rationalizable action profile in the complete information game. \autoref{tab:downward_rationalization_rat_anchor} shows the corresponding iterations of downward rationalizability.
\begin{table}[ht!]
    \centering
    \small
    \caption{Downward rationalizability  for \(p^\prime = (\delta_U, \delta_l)\)}
    \label{tab:downward_rationalization_rat_anchor}
    \renewcommand{\arraystretch}{0.9} 
    \setlength{\tabcolsep}{7pt} 
    \begin{tabular}{c ccccc}
        \toprule
        \multirow{1}{*}{\(\mathsf{D}_{1,p^\prime}^n(\theta_{1,k}), \mathsf{D}_{2,p^\prime}^n(\theta_{2,k}) \)} & \multirow{2}{*}{} & \multicolumn{4}{c}{\(\mathbf{n \rightarrow}\)} \\  
        \cmidrule(lr){3-6}
        \multirow{1}{*}{\(\mathbf{k \downarrow}\)}  & & \textbf{1}  & \textbf{2} & \textbf{3} & \textbf{\(\geq 4\)}   \\ 
        \midrule
        \textbf{1}  & & \myalign{c}{\{U\} , \{l\}} & \myalign{c}{\{U\} , \{l\}} & \myalign{c}{\{U\} , \{l\}} & \myalign{c}{\{U\} , \{l\}} \\
        \textbf{2}  & & \myalign{c}{A_1 , \{l, c\}} & \myalign{c}{\{U\} , \{l\}} & \myalign{c}{\{U\} ,\{l\}} & \myalign{c}{\{U\} , \{l\}}  \\
        \textbf{3}  & & \myalign{c}{A_1 , \{l, c\}} & \myalign{c}{\{U,M\} , \{l, c\}} & \myalign{c}{\{U\} , \{l\}} & \myalign{c}{\{U\} , \{l\}} \\
        \textbf{\(\geq 4\)} & & \myalign{c}{A_1 , \{l, c\}} & \myalign{c}{\{U,M\} , \{l, c\}} & \myalign{c}{\{U,M\} , \{l\}} & \myalign{c}{\{U\}, \{l\}} \\ 
        \bottomrule
    \end{tabular}
\end{table}

As \autoref{prop:monotone} states, one can see that in both \autoref{tab:downward_rationalization} and \autoref{tab:downward_rationalization_rat_anchor}, every column generates two  weakly increasing sequences from top to bottom. \autoref{prop:bounded_behavior} implies that for each type \(\theta_{i,k}\), the implications of higher-order reasoning, seen by differing columns, actually stops at step \(n=k\). Especially, the downward-rationalizable actions for each type are shown on the diagonal of each table. For example, for anchor \(p\), \(\mathsf{D}_{1,p}^\infty(\theta_{1,1}) = \{D\}\), and \(\mathsf{D}_{1,p}^\infty(\theta_{1,2}) = \mathsf{D}_{1,p}^\infty(\theta_{1,3})=\{M,D\}\). Hence we can focus on the entries below and including diagonal of each table. 

Interestingly, in \autoref{tab:downward_rationalization}, the diagonal where \(k=n\) does not respect rationalizability in the complete information game. For example, \(D \in \mathsf{D}_{1,p}^2(\theta_{1,2})\) but \(D\) does not survive the second step in the iterated elimination of dominated actions. Indeed, in the latter, \(D\) is eliminated because action \(r\), to which \(D\) is the best response, is eliminated in step 1; however, for downward rationalization, \(D\) is rationalizable for \(\theta_{1,2}\) because type \(\theta_{1,2}\) could believe that player 2 is a level-\(0\) type. This illustrates an important property of downward rationalizability: a type with higher level can ``replicate'' the belief of the types with lower levels. Hence if an action is downward rationalizable for a level \(k\), so it is for every level \(t \geq k\). The proof of \autoref{prop:monotone} is directly formalizing this intuition.

In contrast, in \autoref{tab:downward_rationalization_rat_anchor}, because the anchor is the Nash equilibrium,  the diagonal respects rationalizability.
\end{example}


\autoref{example: downward-rationalization} illustrate challenges in achieving a general characterization using the well-established solution concepts for the complete information game such as rationalizability, the following result---a benchmark result---provides a \emph{robust} characterization of downward rationalizability across all anchors, but also across all level-\(k\) types. Our notion of robustness can be understood from the perspective of an outside observer who does not know the anchor but seeks to identify all actions that may arise under the maintained assumptions for any possible anchor. Formally, this corresponds to taking the \emph{union} across all anchors, i.e., an existential robustness notion. In this sense, our approach parallels the literature on informational robustness and implementation in robust mechanism design. Moreover, while we adopt this union-based notion throughout, once this robustness across anchors is imposed, our result shows that the choice of robustness across types---\emph{union} vs.\ \emph{intersection}---is immaterial: both notions coincide.\footnote{We do not wish to emphasize this last point too strongly. It is conceptually somewhat unappealing because it mixes the two distinct notions of union-robustness and intersection-robustness.}

\begin{proposition}\label{theorem:characterization_drat}
    For every \(i \in I\),
 \[\rat_i^1 = 
        \bigcup_{p \in \Delta(A_1) \times \Delta(A_2)} 
        \bigcap_{k \in \mathbb{N}}
        \mathsf{D}_{i,p}^\infty(\theta_{i,k})
        =
        \bigcup_{p \in \Delta(A_1) \times \Delta(A_2)} 
        \bigcup_{k \in \mathbb{N}}
        \mathsf{D}_{i,p}^\infty(\theta_{i,k}).
        \]
\end{proposition}

\begin{proof}
Fix \(i \in I\), we will establish
    \begin{enumerate}[nosep]
        \item \(\rat_i^1 \subseteq 
        \bigcup_{p \in \Delta(A_1) \times \Delta(A_2)} 
        \bigcap_{k \in \mathbb{N}}
        \mathsf{D}_{i,p}^\infty(\theta_{i,k})\), and
        \item \(
        \bigcup_{p \in \Delta(A_1) \times \Delta(A_2)} 
        \bigcup_{k \in \mathbb{N}}
        \mathsf{D}_{i,p}^\infty(\theta_{i,k})
         \subseteq \rat_i^1
        \),
    \end{enumerate}
which implies the statement of the theorem because
\[
    \rat_i^1 \overset{\text{by }(1)}{\subseteq} 
    \bigcup_{p \in \Delta(A_1) \times \Delta(A_2)} 
        \bigcap_{k \in \mathbb{N}}
        \rat_{i,p}^\infty(\theta_{i,k}) \subseteq \bigcup_{p \in \Delta(A_1) \times \Delta(A_2)} 
        \bigcup_{k \in \mathbb{N}}
        \rat_{i,p}^\infty(\theta_{i,k})
        \overset{\text{by }(2)}{\subseteq}  \rat_i^1.
\]
\begin{enumerate}[nosep]
    \item Suppose \(a_i \in \rat_i^1\), so that there exists a \(G\)-conjecture \(\nu^i\) such that \(a_i \in r_i(\nu^i)\). Then, take an anchor with \(p_{-i}=\nu^i\) and \(p_i\) arbitrary. Clearly, \(a_i \in 
        \mathsf{D}_{i,p}^1(\theta_{i,1})\). Then, by \autoref{prop:bounded_behavior}, \(a_i \in 
        \mathsf{D}_{i,p}^\infty(\theta_{i,1})\). Furthermore, by \autoref{prop:monotone}, \(\mathsf{D}_{i,p}^\infty(\theta_{i,1}) \subseteq \mathsf{D}_{i,p}^\infty(\theta_{i,k})\) holds for every \(k \in \mathbb{N}\), and the conclusion follows. 
    \item Suppose \(a_i \in \mathsf{D}^\infty_{i,p}(\theta_{i,k})\) for some \(p\) and some \(k\). Then, by definition, \(a_i\) is a best reply to some \(G\)-conjecture.
\end{enumerate}
\end{proof}

In our framework, \autoref{theorem:characterization_drat} is relatively straightforward but showcases the tractability of our approach. It establishes that for all and only justifiable actions, there exists some anchor \(p\) such that the action remains consistent with downward rationalizability across all types. Since justifiable actions are those that survive the first round of iterated elimination of dominated actions---that is, actions consistent with rationality (but not necessarily with higher-order belief in rationality)---in the complete information game, this provides an important insight into behavioral robustness. Therefore, from a market and mechanism design perspective, \autoref{theorem:characterization_drat} offers a novel argument for the use of \emph{simple} mechanisms: If the designer seeks to implement a mechanism that is \emph{robust} to (as-if) \emph{level-\(k\) reasoning} and also resilient to the assumed \emph{anchor}, then the mechanism should rely solely on players' rationality and not on any (higher-order) reasoning. 

In contrast\footnote{We thank Antonio Penta for urging us to be more explicit about this connection.} to our method, which embeds the complete-information game into one with payoff uncertainty, \citet{strzalecki2014depth} directly models uncertainty about co-players' levels using \citeauthor{harsanyi1967games}'s (\citeyear{harsanyi1967games}) type structures, which he calls \emph{cognitive type spaces}.\footnote{\citet[Definition 4]{alaoui2016endogenous} consider a more general notion of cognitive type spaces. To be clear, what follows concerns \citeauthor{strzalecki2014depth}'s notion, not the broader concept of \citet{alaoui2016endogenous}.}  In abstract, but standard, game-theoretic terminology, his approach thus differs from ours in the same way that Bayesian games differ from games with payoff uncertainty.\footnote{For the distinction in the abstract, game-theoretic setting, see the introductions of \citet{battigalli1999rationalizability, battigalli2003rationalizability} and \citet{battigalli2003rationalization}.} Both approaches, however, do allow for hetergenous perception about their co-players levels.\footnote{\citet*{alaoui2025coordination} illustrate how perceived heterogeneity in \emph{cognitive abilities} can help players coordinate.}

With this comparison in mind, it is well known that in the abstract, game-theoretic setting these two approaches are related via \emph{informational robustness}, as studied by \citet{bergemann2016bayes, bergemann2017belief}.\footnote{Early results in the same spirit include \citet{aumann1987correlated}, \citet{brandenburger1987rationalizability}, \citet{forges1993five, forges2006correlated}, \citet{battigalli2003rationalization}, and \citet{dekel2007interim}.} We conjecture that downward rationalizability admits a similar informational robustness interpretation when one starts from \citeauthor{strzalecki2014depth}'s cognitive equilibrium and considers robustness across all cognitive type spaces.\footnote{\citet{strzalecki2014depth} also considers cognitive rationalizability.  A similar informational robustness foundation conjecture holds in that case as well. See \citet{ziegler2022informational} for the corresponding result in the abstract, game-theoretic setting.}

\section[Level-k Rationalizability]{Level-\(k\) Rationalizability}\label{sec:L-rat}
As seen in the previous section, even when fixing a particular anchor \( p \), downward rationalizability differs from the classic models of bounded and iterated reasoning in games. Indeed, we will show next that all the established level-\(k\) models mentioned in \autoref{sec:complete_info} are special cases of our framework using the game with payoff uncertainty. The two classic models can be derived by imposing more restrictive (and transparent) assumptions on first-order beliefs.

For the classic level-\(k\) solution, it is intuitively clear that (K2) is too weak because it allows a given type to deem possible any lower types. In contrast, the implicit behavioral assumption in the classic level-\(k\) model is that a given \(k\)-type considers only \((k-1)\)-types of their co-player as possible, disregarding even lower types. To capture the classic level-\(k\) solution within our \(\Delta\)-rationalizability framework for the game with payoff uncertainty start with (K1) and (K2) as introduced above, and add the following restriction for a type \(\theta_{i,k}\):
\begin{enumerate}[nosep]
	\item[\(\textup{(KL)}\).] \( \supp \marg_{\Theta_{-i}} \mu^{\theta_{i,k}} \subseteq \{ \theta_{-i,k-1}\}, \)
\end{enumerate}
which, like (K2), is a restriction on exogenous beliefs only.

The resulting belief restrictions give rise to a version of \(\Delta\)-rationalizability, which we will refer to as \emph{\(\mathsf{L}\)-rationalizable} (short for \emph{level-\(k\) rationalizable}) action-type combinations (under anchor \(p\)), denoted by \(\mathsf{L}_p^{\infty}:=\mathsf{L}_{1,p}^{\infty} \times \mathsf{L}_{2,p}^{\infty}\). We will also use the same notation for other related and derived objects. Note that Condition (KL) is stronger than Condition (K2). However, we choose to state it as an independent condition because we aim to preserve the general framework defined by (K1) and (K2) while exploring how this framework can provide a foundation for various solution concepts of level-\(k\) reasoning by incorporating appropriate additional conditions. By imposing more restrictive yet transparent belief restrictions, we immediately obtain that \(\mathsf{L}\)-rationalizability \emph{refines} downward rationalizability, because it is well known that more restrictive belief assumptions lead to refinements \citep[Remark 22]{battigalli2023game}.

\begin{remark}\label{remark:level-k-subset}
	For every \(i \in I\) and every \(n \in \mathbb{N}_0 \cup \{\infty\}\),
	\(
	\mathsf{L}_{i,p}^{n} \subseteq \mathsf{R}_{i,p}^{n}.
	\)
\end{remark}

Furthermore, as with downward rationalizability, it holds for \(\mathsf{L}\)-rationalizability that a \(k\)-type behaves \emph{as if} they reason at most \(k\) steps. Thus, the explanation given just before \autoref{prop:bounded_behavior} applies here verbatim as well.

\begin{remark}\label{coro:level-k-bounded-reason}
	For every \(i \in I\), every \( k \in \mathbb{N} \), and every \( n \geq k \),  
	\(
	\mathsf{L}^n_{i, p}(\theta_{i,k}) = \mathsf{L}^k_{i, p}(\theta_{i,k}).
	\)
\end{remark}

A key difference from the coarser downward rationalizability is that the monotonicity result of \autoref{prop:monotone} does not carry over to \(\mathsf{L}\)-rationalizability . Given the formal connection with the classic level-\(k\) solution, established later in \autoref{theorem:coincidence-classic}, this is well known in the literature. 

Another well-known fact is that ``[classic level-]\(k\) respects \(k\)-rationalizability'' \citep[p.14]{crawford2013structural}, which remains true in our payoff uncertainty framework as well. In fact, we establish such a result for \(\mathsf{L}\)-rationalizability, which holds for all \(k \geq n\). Together with \autoref{theorem:coincidence-classic} below, the literal interpretation of the aforementioned quote could suggest that the statement applies only to the case \(n = k\).\footnote{\citet[Proposition 1]{schipper2024level} provides a formal proof for the \(n = k\) case and, therefore, our result generalizes theirs.}

\begin{proposition}[Respecting \(n\)-rationalizability]\label{prop:classic_k_respects_rat}
   Fix an anchor \(p\). For every \(n \in \mathbb{N}_0\), \(k \geq n\), and \(i \in I\),
\(
	\mathsf{L}_{i,p}^n(\theta_{i,k}) \subseteq \rat_i^n.
\)
\end{proposition}

\begin{proof}
    Fix an anchor \(p\). We prove the statement by induction on \(n\). The base case with \(n=0\) is trivial, since by definition, for every \(k \geq 0\) and every \(i \in I\),
\(
	\mathsf{L}_{i,p}^0(\theta_{i,k}) = A_i = \rat_i^0.
\)
Now, assume the statement holds for some \(n > 0\), and consider \(k \geq n+1\), \(i \in I\), and \(a_i \in L_{i,p}^{n+1}(\theta_{i,k})\). Then, there exists a \(\hat{G}\)-conjecture \(\mu^i\) such that \(a_i \in r_i(\marg_{A_{-i}} \mu^i)\) and 
\(
\mu^i\left(\{\theta_{-i,k-1}\} \times \mathsf{L}_{-i,p}^{n}(\theta_{-i,k-1})\right) = 1.
\)
Since \(k-1\geq n\), by the induction hypothesis, we have \(\mathsf{L}_{-i,p}^{n}(\theta_{-i,k-1}) \subseteq \rat_{-i}^n\). Thus, defining \(\nu^i := \marg_{A_{-i}} \mu^i\) ensures that \(\nu^i(\rat_{-i}^n) = 1\). Therefore, \(a_i \in \rat_i^{n+1}\).
\end{proof}

The following statement---the main result of this section---shows that the classic level-\(k\) solution is tightly connected to \(\mathsf{L}\)-rationalizability , which itself is characterized by the behavioral implications of rationality, common belief in rationality, and the transparency of (K1), (K2), and (KL).

\begin{theorem}\label{theorem:coincidence-classic}
Fix an anchor \(p\). For every \(k \in \mathbb{N}\), \(n \in \mathbb{N}_0\), and every \(i \in I\),
\[
	\mathsf{L}_{i,p}^n(\theta_{i,k}) =
	\begin{cases}
		R_i^n, & \text{if } n < k, \\
		L_i^k[p], & \text{if } n \geq k.
	\end{cases}
\]
\end{theorem}

This theorem shows that, for each type \(k\), \(\mathsf{L}\)-rationalizability s the complete-information rationalizability procedure up to reasoning order \(k-1\), and at order \(k\) yield player \(i\)'s classic level-\(k\) behavior. In particular, when no player has dominated actions, \(\mathsf{L}\)-rationalizability reproduces the classic level-\(k\) model not only in terms of behavioral predictions but also in terms of the underlying reasoning process. Specifically, for a type \(k\), no action is eliminated before order \(k\), and at order \(k\) the predicted behavior corresponds to the classic level-\(k\) prediction. Thus, the ``diagonal'' of \(\mathsf{L}\)-rationalizability coincides exactly with classic level-\(k\) behavior.

Together with the epistemic foundation of \(\Delta\)-rationalizability \citep{battigalli2007interactive,battigalli2013transparent}, the theorem yields the following nuanced epistemic foundation of the classic level-\(k\) model. We state it informally, since spelling out all objects here would take us too far afield and follows standard arguments.

\begin{corollary}[Epistemic foundation of classic level-\(k\), informal]
Fix a finite order \(n\geq 1\) of higher-order reasoning about rationality, and consider a complete epistemic type structure.\footnote{More precisely, we require completeness relative to the first-order belief restrictions \(\textup{(K1)}\) and \(\textup{(KL)}\), meaning that the type structure contains all belief hierarchies consistent with these restrictions.} Then transparency of \(\textup{(K1)}\),  \(\textup{(K2)}\), and \(\textup{(KL)}\), together with rationality and \((n-1)\)-fold belief in rationality, yield behavioral implications coinciding with the set of \(n\)-rationalizable actions for a type \(k\) with \(k>n\), and the classic level-\(k\) actions for a type \(k\) with \(k \le n\).
\end{corollary}

\begin{proof}[Proof of \autoref{theorem:coincidence-classic}]
We prove the claim by induction on \(k \in \mathbb{N}\).

\noindent \textbf{Base Case (\(k=1\)):}  
We consider three cases.
\begin{enumerate}[leftmargin=1.5\parindent, align=left, nosep]
    \item[\textbf{Case B.1} (\(n < 1\)):]  
    Since \(n \in \mathbb{N}_0\), we have \(n=0\). By definition, for every \(i \in I\),
    \(
        \mathsf{L}_{i,p}^0(\theta_{i,1}) = A_i = \rat_i^0.
    \)

    \item[\textbf{Case B.2} (\(n = 1\)):]  
    The result follows directly from conditions (K1) and (K2).

    \item[\textbf{Case B.3} (\(n > 1\)):]  
    The result follows from \autoref{coro:level-k-bounded-reason} together with the previous case.
\end{enumerate}

\noindent \textbf{Inductive Step:}  
Assume that the statement holds for some \(k > 1\); that is, for every \(n \in \mathbb{N}_0\) and every \(i \in I\),
\[
    \mathsf{L}_{i,p}^n(\theta_{i,k}) =
    \begin{cases}
        \rat_i^n, & \text{if } n < k, \\
        L_i^k[p], & \text{if } n \geq k.
    \end{cases}
\]
We now prove the statement for \(k+1\). Again, we distinguish three cases.
\begin{enumerate}[leftmargin=1.5\parindent, align=left, nosep]
    \item[\textbf{Case I.1} (\(n < k+1\)):]  
    If \(n=0\), the argument from Case B.1 applies. Suppose instead that \(n \geq 1\). By the induction hypothesis, for all \(j \in I\) and all \(m < k\) (in particular, for \(m = n-1\)),
    \[
        \{\theta_{j,k}\} \times \mathsf{L}_{j,p}^{m}(\theta_{j,k}) = \{\theta_{j,k}\} \times \rat_j^{m}.
    \]
    Now, consider \(i \in I\) and suppose \(a_i \in \rat_i^n\). Then there exists \(\nu^i \in \Delta(A_{-i})\) such that
    \(
        a_i \in r_i(\nu^i) \) and \(\nu^i\bigl(\rat_{-i}^{n-1}\bigr) = 1.
    \)
    Define the \(\hat{G}\)-conjecture \(\mu^i\) by
    \[
        \mu^i(a_{-i},\theta_{-i}) :=
        \begin{cases}
            \nu^i(a_{-i}), & \text{if } \theta_{-i} = \theta_{-i,k}, \\[1mm]
            0, & \text{otherwise.}
        \end{cases}
    \]
    It is straightforward to verify that \(\mu^i\) satisfies conditions (K1), (K2), and (KL). By construction and the induction hypothesis,
    \(
        \mu^i\Bigl(\{\theta_{-i,k}\} \times \mathsf{L}_{-i,p}^{n-1}(\theta_{-i,k})\Bigr) = 1.
    \)
    Hence, \(a_i \in \mathsf{L}_{i,p}^n(\theta_{i,k+1})\).

    Note that the other inclusion holds automatically by \autoref{prop:classic_k_respects_rat}.

    \item[\textbf{Case I.2} (\(n = k+1\)):]  
    Consider \(i \in I\) and first suppose that \(a_i \in \mathsf{L}_{i,p}^{k+1}(\theta_{i,k+1})\). Then there exists a justifying \(\hat{G}\)-conjecture \(\mu^i\) such that
    \(
        \mu^i\Bigl(\{\theta_{-i,k}\} \times \mathsf{L}_{-i,p}^{k}(\theta_{-i,k})\Bigr) = 1.
    \)
    By the induction hypothesis, we have \(\mathsf{L}_{-i,p}^{k}(\theta_{-i,k}) = \lk[-i]^k\), so that
    \(
        \mu^i\Bigl(\{\theta_{-i,k}\} \times \lk[-i]^k\Bigr) = 1.
    \)
    Taking \(\nu^i := \marg_{A_{-i}}\mu^i\) then shows that \(a_i \in \lk[i]^{k+1}\).

    Conversely, suppose that \(a_i \in \lk[i]^{k+1}\). Then there exists \(\nu^i \in \Delta(A_{-i})\) such that
    \(
        a_i \in r_i(\nu^i) \) and \( \nu^i\bigl(\lk[-i]^{k}\bigr) = 1.
    \)
    Define the \(\hat{G}\)-conjecture \(\mu^i\) as before by
    \[
        \mu^i(a_{-i},\theta_{-i}) :=
        \begin{cases}
            \nu^i(a_{-i}), & \text{if } \theta_{-i} = \theta_{-i,k}, \\[1mm]
            0, & \text{otherwise.}
        \end{cases}
    \]
    It follows that \(\mu^i\) satisfies conditions (K1), (K2), and (KL). By the induction hypothesis,
    \(
        \mu^i\Bigl(\{\theta_{-i,k}\} \times \mathsf{L}_{-i,p}^{k}(\theta_{-i,k})\Bigr) = 1,
    \)
    so that \(a_i \in \mathsf{L}_{i,p}^{k+1}(\theta_{i,k+1})\).

    \item[\textbf{Case I.3} (\(n > k+1\)):]  
    The conclusion in this case follows from the previous case together with \autoref{coro:level-k-bounded-reason}.
\end{enumerate}

Combining, the inductive step is complete, and the statement follows.
\end{proof}

Since, in our finite setting, both rationalizability and the classic level-\(k\) solution are always non-empty for the complete information game, we immediately obtain from \autoref{theorem:coincidence-classic} that for every \(i \in I\) and every \(k \in \mathbb{N}_0\), \(\mathsf{L}^n_{i, p}(\theta_{i,k}) \neq \emptyset\) for each \(n \in \mathbb{N}_0\), and \(\mathsf{L}^\infty_{i, p}(\theta_{i,k}) \neq \emptyset\). This observation, together with \autoref{remark:level-k-subset}, also establishes that downward rationalizability is always well-defined in the sense of yielding non-empty predictions.\footnote{It is well known---e.g., \citet[Remark 3.2]{battigalli2003rationalization} or \citet[Theorem 31]{battigalli2023game}---that for restrictions on exogenous beliefs, i.e. those concerning only \(\theta_{-i}\), \(\Delta\)-rationalizability guarantees non-empty predictions. Here, however, our restrictions are not limited to such cases, because of (K1). Thus, we cannot rely directly on this usual argument.}

\autoref{theorem:coincidence-classic} has a direct empirical interpretation. In our framework, \(k\) summarizes a belief-driven component, while \(n\) captures cognitive depth. Depending on the context, either or both may vary with the environment (e.g., feedback about opponents, complexity, cognitive load, or time pressure). The theorem implies that when \(n<k\) behavior is disciplined by \(n\)-step rationalizability, whereas when \(n\ge k\) it coincides with the classic level-\(k\) prediction under the given anchor. Hence, a subject with stable \(k\) but varying \(n\) may be associated by different ``levels'' across games in a standard exercise, even though her behavior is coherent with the fixed $k$-type once the two components are separated. This distinction can be probed by combining belief data or belief-control designs with within-subject variation in deliberation conditions \citep[e.g.,][]{chen2025measuring}.

The next example illustrates \autoref{theorem:coincidence-classic}.

\begin{example}[continued from \autoref{example: downward-rationalization}]\label{example:L-rationalization}
We still consider the game in \autoref{tab:iterated_elimination_example}. Recall the rationalizability and classic level-\(k\) behavior in this complete information game:
\begin{align*}
    \rat^1 &= A_1 \times \{l, c\}, 
    &\quad&
    &\lk^1 & = \{D\} \times \{c\},\\[-0.33em]
    \rat^2 &= \{U,M\} \times \{l, c\}, 
    &\quad&
    &\lk^2 & = \{M\} \times \{c\},\\[-0.33em]
    \rat^3 &= \{U,M\}\times \{l\}, 
    &\quad&
    &\lk^3 & = \{M\} \times \{l\},\\[-0.33em]
    \rat^4 &= \{U\} \times \{l\} = \rat^\infty, \text{ and} 
    &\quad&
    &\lk^4 & = \{U\} \times \{l\}.
\end{align*}
\(\mathsf{L}\)-rationalizability is shown in \autoref{tab:L_rationalization}.

\begin{table}[ht!]
    \centering
    \small
    \caption{\(\mathsf{L}\)-rationalizability with \(p=(\delta_D, \delta_r).\)}
    \label{tab:L_rationalization}
    \renewcommand{\arraystretch}{0.9} 
    \setlength{\tabcolsep}{7pt} 
    \begin{tabular}{c ccccc}
        \toprule
        \multirow{1}{*}{\(\mathsf{L}_1^n(\theta_{1,k}), \mathsf{L}_2^n(\theta_{2,k}) \)} & \multirow{2}{*}{} & \multicolumn{4}{c}{\(\mathbf{n \rightarrow}\)} \\  
        \cmidrule(lr){3-6}
        \multirow{1}{*}{\(\mathbf{k \downarrow}\)}  & & \textbf{1}  & \textbf{2} & \textbf{3} & \textbf{\(\geq 4\)}   \\ 
        \midrule
        \textbf{1}  & & \( \{D\}, \{c\} \) & \( \{D\}, \{c\} \) & \( \{D\}, \{c\} \) & \( \{D\}, \{c\} \) \\
        \textbf{2}  & & \( A_1, \{l, c\} \) & \( \{M\}, \{c\} \) & \( \{M\}, \{c\} \) & \( \{M\}, \{c\} \)  \\
        \textbf{3}  & & \( A_1, \{l, c\} \) & \( \{U, M\}, \{l, c\} \) & \( \{M\}, \{l\} \) & \( \{M\}, \{l\} \) \\
        \textbf{\(\geq 4\)} & & \( A_1, \{l, c\} \) & \( \{U, M\}, \{l, c\} \) & \( \{U, M\}, \{l\} \) & \( \{U\}, \{l\} \) \\  
        \bottomrule
    \end{tabular}
\end{table}

In accordance with \autoref{theorem:coincidence-classic}, in \(\mathsf{L}\)-rationalizability, at each reasoning step \(n\), the actions associated with type-\(k\) coincides either with rationalizability in the complete-information game or with the classic level-\(k\) behavior. The former happens when the reasoning step (\(n\)) does not reach the type level (\(k\)), and the latter occurs when the reasoning reaches and exceeds this level.
\end{example}

\autoref{example:L-rationalization}, and in particular the last row in \autoref{tab:L_rationalization}, reveals an interesting implication of \autoref{theorem:coincidence-classic}. 

\begin{corollary}\label{corollary: indifference}
If the complete information game \(G\) is dominance solvable in \(n^*\) steps (i.e., the iterated elimination of dominated actions stops at step \(n^*\) and each \(R^{n^*}_i\) is a singleton), then for every anchor \(p\), every \(n \in \mathbb{N}\), every \(k \geq n^*\), and every \(i \in I\),  \(\mathsf{L}_{i,p}^n(\theta_{i,k}) = R_i^n\).
\end{corollary}

This corollary formally states that in a dominance-solvable game requiring \(n^*\) steps, if \(k\) is sufficiently large, the anchor becomes irrelevant under \(\mathsf{L}\)-rationalizability. The reason is that, as \autoref{theorem:coincidence-classic} shows, the anchor \(p\) can only affect the reasoning of a \(k\)-type at step \(k\) or later. If \(k \geq n^*\), however, the standard rationalization process has already eliminated all other actions before the anchor can matter. Thus, the anchor has no effect on the outcome, regardless of its specification.

This also clarifies the sense in which level-\(k\) reasoning remains consistent with standard complete-information rationalizability. Once the type index \(k\) and the reasoning order \(n\) are allowed to vary independently, a given \(k\)-type need not in general behave in a way that is consistent with \(n\) rounds of rationalizability. However, for every order of reasoning \(n\), there are always types whose behavior is consistent with \(R_i^n\), and taking the union of \(\mathsf{L}\)-rationalizable actions across those consistent types recovers exactly \(R_i^n\). In this sense, the incomplete-information construction remains aligned with standard complete-information reasoning. It also helps clarify the connection to \citet{brandenburger2020two}, because their complete-information epistemic model and our payoff-uncertainty model use different primitives and different mechanisms, but both preserve a close link between level-\(k\) behavior and \(n\)-rationalizability once one conditions on the appropriate types. The formal statements, proofs, and a more detailed comparison with \citet{brandenburger2020two} are collected in Appendix~\ref{app:sec:consistency}.

\subsection{Robustness}\label{subsec:Lk_robust}
Finally, we explore an analogue of the robustness question previously examined for downward rationalizability. Given \autoref{theorem:coincidence-classic}, it is clear that \(\mathsf{L}_{i,p}^\infty(\theta_{i,k}) = \lk[i]^k \subseteq R_i^k\) holds for any \(k \in \mathbb{R}\)---but more importantly, also for any anchor \(p\). That is, \(\bigcup_p \mathsf{L}_{i,p}^\infty(\theta_{i,k}) \subseteq R_i^k\). In other words, \(\mathsf{L}\)-rationalizability does respect \(k\)-rationalizability robustly. But does the converse hold? Does \(k\)-rationalizability respect robust \(\mathsf{L}\)-rationalizability, i.e., \(R_i^k \subseteq \bigcup_p \mathsf{L}_{i,p}^\infty(\theta_{i,k})\)?

Perhaps somewhat surprisingly, the answer is \emph{no}, and we illustrate this next with an exemplary game in which the inclusion fails for \(k=2\)---the smallest value of \(k\) for which it can fail.\footnote{We thank Marciano Siniscalchi for encouraging us to explore this question. After developing our example, we discovered that \citet{brandenburger2020two} had already constructed an example that---among other things---also illustrates the failure of the inclusion. Their example is more general, as it exhibits an action that is rationalizable but not level-\(k\) rationalizable for any \(k \geq 2\) and any anchor. Since our aim is only to illustrate a failure of the inclusion for a given \(k\), we consider a \(3 \times 3\) game, rather than the \(4 \times 4\) game used in their construction.}

\begin{example}[\(2\)-rationalizable action is not robustly level-\(k\) rationalizable for a level-\(2\) type] Consider the game in \autoref{tab:Lk_folk_theorem}. 

    \begin{table}[ht!]
    \centering
    \small
    \caption{A game illustrating \(\bigcup_p \mathsf{L}_{i,p}^\infty(\theta_{2,2}) \subsetneq R_2^2\)}
    \label{tab:Lk_folk_theorem}
    \renewcommand{\arraystretch}{0.75} 
    \setlength{\tabcolsep}{10pt} 
    \begin{tabular}{c ccc}
        \toprule
        Player 1 \textbackslash \, 2 & \textbf{\(l\)}  & \textbf{\(c\)} & \textbf{\(r\)}  \\ 
        \midrule
        \textbf{\(U\)}  & \(6, 3\) & \(0, 6\) & \(0, 0\)  \\
        \textbf{\(M\)}  & \(0, 3\) & \(6, 0\) & \(0, 6\)  \\
        \textbf{\(D\)}  & \(4, 0\) & \(4, 6\) & \(4, 6\)  \\ 
        \bottomrule
    \end{tabular}
\end{table}

\autoref{fig:anti-folk} shows the simplex of conjectures for both players. Since anchors are the same mathematical objects, these simplices also represent the anchors (with player indices swapped). Moreover, they depict the exact best-reply sets (EBRS) and their corresponding sets of justifiable conjectures---those conjectures that render the indicated actions exactly the best replies.\footnote{We adopt the terminology of \citet{weinstein2020best}.}

In \autoref{fig:anti-folk}, each number in a circle indicates a region in the set of each \(i\)'s anchors, and the associated action(s) is(are) the best response(s) to all the anchors in that area. For example, ``\Circled{1} \(\{U\}\)'' indicates that if player 2's anchor assigns sufficiently high probability to \(l\), then \(U\) is uniquely optimal. For all anchors with \(p_2(l)=1/3\) (left blue line in the left panel), both \(U\) and \(D\) are optimal, indicated by ``\Circled{2} \(\{U,D\}\)''. In the right panel, focused on player 2, we see ``\Circled{2} \(\{l,c,r\}\).'' There is a unique anchor of player 1 that makes \(l\) part of an exact best-reply set: the fifty-fifty mixture between \(U\) and \(M\), as indicated by \Circled{2} in \autoref{fig:anti-folk} (b). Together, these two panels cover all possible level-\(1\) \(\mathsf{L}\)-rationalizable actions across all anchors.

\begin{figure}
  \centering
  \newlength{\triangleSide}
  \setlength{\triangleSide}{5.65cm} 
  \pgfmathsetlengthmacro{\triangleHeight}{\triangleSide*sqrt(3)/2}
  \begin{subfigure}[b]{0.48\textwidth}
    \centering
    \begin{tikzpicture}[scale=1]
      \coordinate (A) at (0,0);
      \coordinate (B) at (\triangleSide,0);
      \coordinate (C) at (0.5\triangleSide,\triangleHeight);
      \draw[thick] (A) -- (B) -- (C) -- cycle;
      \node[below left]  at (A) {\(\delta_l\)};
      \node[below right] at (B) {\(\delta_r\)};
      \node[above]       at (C) {\(\delta_c\)};
      \coordinate (D) at ($(A)!1/3!(B)$);
      \coordinate (E) at ($(A)!1/3!(C)$);
      \draw[blue, thick] (D) -- (E);
      \coordinate (F) at ($(A)!2/3!(B)$);
      \coordinate (G) at ($(B)!1/3!(C)$);
      \draw[blue, thick] (F) -- (G);
      \coordinate (R1) at ($0.333*(A)+0.333*(D)+0.333*(E)+(0.03,0)$); 
      \coordinate (R2) at ({0.5*\triangleSide},{0.6*\triangleHeight}); 
      \coordinate (R3) at ($0.333*(B)+0.333*(F)+0.333*(G)+(0.03,0)$); 
      \node[anchor=center] at (R1) {\footnotesize \Circled{1} \(\{U\}\)};
      \node[anchor=center] at (R2) {\footnotesize \Circled{3} \(\{D\}\)};
      \node[anchor=center] at (R3) {\footnotesize \Circled{5} \(\{M\}\)};
      \node at ($(D)!0.7!(E)+(0.575,0.0)$) {\footnotesize \Circled[fill color=white]{2} \(\{U,D\}\)}; 
      \node at ($(F)!0.7!(G)-(0.575,0.0)$) {\footnotesize  \(\{M,D\}\)\Circled[fill color=white]{4} };  
    \end{tikzpicture}
    \caption{Player 1}
  \end{subfigure}
  \quad
  \begin{subfigure}[b]{0.48\textwidth}
    \centering
    \begin{tikzpicture}[scale=1]
      \coordinate (A) at (0,0);
      \coordinate (B) at (\triangleSide,0);
      \coordinate (C) at (0.5\triangleSide,\triangleHeight);
      \draw[thick] (A) -- (B) -- (C) -- cycle;
      \node[below left]  at (A) {\(\delta_U\)};
      \node[below right] at (B) {\(\delta_D\)};
      \node[above]       at (C) {\(\delta_M\)};
      \coordinate (D) at ($(A)!1/2!(C)$);
      \draw[blue, thick] (D) -- (B);
      \fill[blue] (D) circle (3pt);
      \coordinate (R1) at ({0.33*\triangleSide},{0.2*\triangleHeight});
      \coordinate (R2) at ({0.5*\triangleSide},{0.7*\triangleHeight});
      \node[anchor=center] at (R1) {\footnotesize \Circled{1} \(\{c\}\)};
      \node[anchor=center] at (R2) {\footnotesize \Circled{3} \(\{r\}\)};
      \node[above left] (L) at ($(D)-(0.0, 0.2)$) {\footnotesize \(\{l,c,r\}\)\Circled{2}};
      \node (R) at ($(D)!0.5!(B)+(0.0,0.26)$) {\footnotesize \Circled[fill color=white]{4} \(\{r,c\}\)};
    \end{tikzpicture}
    \caption{Player 2}
  \end{subfigure}
  \caption{EBRS and the corresponding conjectures/anchors} \label{fig:anti-folk}
\end{figure}
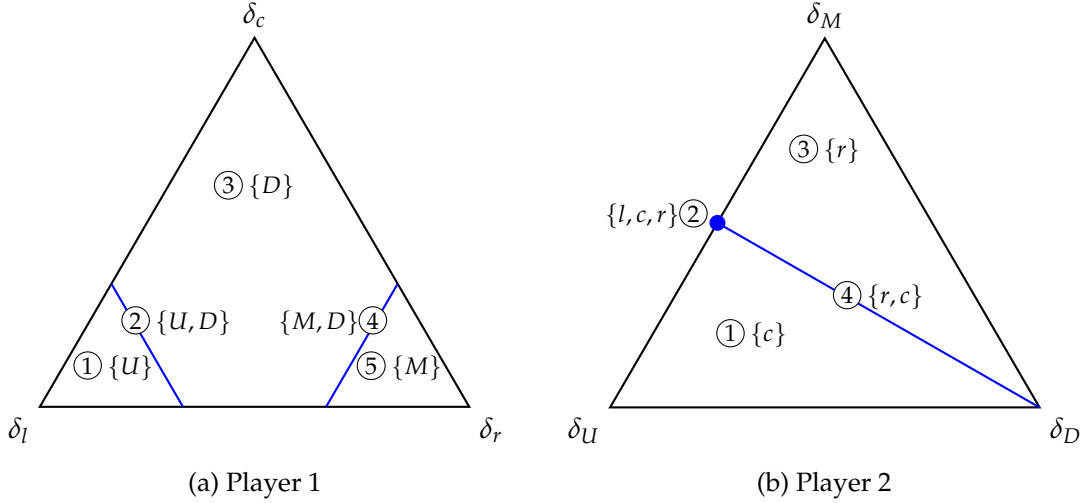

Now, we can use these EBRS to obtain all possible level-\(2\) \(\mathsf{L}\)-rationalizable actions across all anchors. For player 2, this is summarized in \autoref{tab:antifolk_level2}. For example, if player 2's anchor is in region \Circled{1}, then the corresponding Player 1 EBRS is \(\{U\}\). In that case, player 2's exact EBRS---given conjectures concentrated on \(\{U\}\)---is \(\{c\}\). If the anchor lies within region \Circled{2}, then the relevant player 1 EBRS is \(\{U,D\}\). For conjectures concentrated on this set, this yields two EBRS for player 2: \(\{c,r\}\) and \(\{c\}\). From this table, we see that \(l\) does not appear. Therefore, \(l\) cannot be level-\(2\) \(\mathsf{L}\)-rationalizable. However, all actions of all players---and \(l\), in particular---are rationalizable (\emph{a fortiori}, \(2\)-rationalizable) in the complete information game.

\begin{table}[ht!]
    \centering
    \small
    \caption{Level-\(2\) behavior corresponding to anchors}
    \label{tab:antifolk_level2}
    \renewcommand{\arraystretch}{0.75} 
    \setlength{\tabcolsep}{15pt} 
    \begin{tabular}{c cccc}
        \toprule
        & \multicolumn{4}{c}{Level-\(2\) behavior of Player 2}  \\ 
        Player 1 EBRS & EBRSs & \(\bigcup\)EBRSs  \\ 
        \midrule
        \Circled{1} \(\{U\}\) & \(\{c\}\)& \(\{c\}\)\\
        \Circled{2} \(\{U,D\}\)& \(\{c,r\}, \{c\}\)& \(\{c,r\}\) \\
        \Circled{3} \(\{D\}\)& \(\{c,r\}\)& \(\{c,r\}\) \\
        \Circled{4} \(\{M,D\}\) & \(\{c,r\}, \{c\}\)& \(\{c,r\}\) \\ 
        \Circled{5} \(\{M\}\)& \(\{r\}\)& \(\{r\}\) \\
        \bottomrule
    \end{tabular}
\end{table}

Why does this discrepancy arise? The key lies in the support of a conjecture that justifies \(l\). Such a conjecture must assign probability only to \(\{U,M\}\). However, no EBRS of player 1 contains \(\{U,M\}\) as a subset. Thus, while in the classic level-\(k\) model the anchor propagates through all levels via the EBRS, this propagation fails for rationalizability in the complete-information game. In the latter case, the justifying conjecture for actions in \(R^1\) is irrelevant and can be ignored. Here, we require only the union of all EBRS. By contrast, the classic level-\(k\) solution across all anchors relies on the entire EBRS collection for the next iteration, not merely its union.
\end{example}


\section{Cognitive-Hierarchy Rationalizability}\label{sec:CH-rat}
Next, we turn to the CH model. To capture the CH solution, we retain conditions (K1) and (K2) and add an additional condition, labeled (KC), to incorporate the (fixed) level distribution \(f\)---a crucial ingredient of the CH model as introduced in \autoref{subsec:static_CH}---into our game with payoff uncertainty framework. Specifically, a conjecture for a given type \(\theta_{i,k}\) must also satisfy the following restriction:
\begin{itemize}[nosep]
	\item[\textbf{(KC)}.] \(\mu^i(\theta_{-i,t}) = f^{k-1}(t) \quad \text{for every } t = 0, \ldots, k-1.\)\footnote{Recall, for \(f \in \Delta_+(\mathbb{N}_0)\) and for every \(k\in \mathbb{N}\), we define \(f^k(t) :=  \frac{f(t)}{\sum_{t'=0}^{k-1}f(t')}\) for every \(t = 0, \ldots ,k-1\).}
\end{itemize}

Similarly to before, we obtain belief restrictions \(\Delta\) satisfying (K1), (K2), and~(KC), and a corresponding version of \(\Delta\)-rationalizability that we refer to as \emph{\(\mathsf{C}\)-rationalizability} (short for \emph{Cognitive-Hierarchy rationalizability}), with notation \(\mathsf{C}^{\infty}_{p,f} := \mathsf{C}_{1,p,f}^{\infty} \times \mathsf{C}_{2,p,f}^{\infty}\). We will also use this notation for other related and derived objects.

Again, because the belief restrictions are more restrictive than those in downward rationalizability, we immediately have the following remark.

\begin{remark}\label{remark:CH-subset}
	For every \(i \in I\) and every \(n \in \mathbb{N}_0 \cup \{\infty\}\),
	\(
	\mathsf{C}_{i,p,f}^{n} \subseteq R_{i}^{n}.
	\)
\end{remark}

Also like before, it holds that a player of type \(\theta_{i,k}\) behaves \emph{as if} they reason at most \(k\) steps.

\begin{remark}\label{CH: bounded_reasoning}
	For every \(i \in I\), every \( k \in \mathbb{N}_0 \) and every \( n \geq k \),  
	\(
	\mathsf{C}^n_{i,p,f}(\theta_{i,k}) = \mathsf{C}^k_{i,p,f}(\theta_{i,k}).
	\)
\end{remark}

We now return to our running example to illustrate how \(\mathsf{C}\)-rationalizability works for a concrete game.

\begin{example}[continued from \autoref{example: downward-rationalization}]\label{example:CH-rationalization}
As before, consider the game in \autoref{tab:iterated_elimination_example} with anchor \(p = (\delta_D, \delta_r)\). Now, we also need to specify a level distribution. For illustrative purposes, we take \(f\) to be a geometric distribution with parameter \(1/2\), so that
\begin{align*}
    f^2(0) &= \tfrac{2}{3}, &\quad f^2(1) &= \tfrac{1}{3}, \\
    f^3(0) &= \tfrac{4}{7}, &\quad f^3(1) &= \tfrac{2}{7}, &\quad f^3(2) &= \tfrac{1}{7}, \\
    f^4(0) &= \tfrac{8}{15}, &\quad f^4(1) &= \tfrac{4}{15}, &\quad f^4(2) &= \tfrac{2}{15}, &\quad f^4(3) &= \tfrac{1}{15}.
\end{align*}
Looking at the CH level-\(k\) solution, we obtain \(CH^1[p,f]=\{(D,c)\}\), \(CH^2[p,f]=\{M,D\}\times \{c\}\), and \(CH^k[p,f]=\{(M,c)\}\) for \(k \geq 3\).
The corresponding \(\mathsf{C}\)-rationalizability process is shown in \autoref{tab:CH_rationalization}.

\begin{table}[ht!]
    \centering
    \small
    \caption{``Geometric'' \(\mathsf{C}\)-rationalizability with \(p=(\delta_D, \delta_r).\)}
    \label{tab:CH_rationalization}
    \renewcommand{\arraystretch}{0.9} 
    \setlength{\tabcolsep}{7pt} 
    \begin{tabular}{c ccccc}
        \toprule
        \multirow{1}{*}{\(\mathsf{C}_{1,p,f}^n(\theta_{1,k}), \mathsf{C}_{2,p,f}^n(\theta_{2,k}) \)} & \multirow{2}{*}{} & \multicolumn{4}{c}{\(\mathbf{n \rightarrow}\)} \\  
        \cmidrule(lr){3-6}
        \multirow{1}{*}{\(\mathbf{k \downarrow}\)}  & & \textbf{1}  & \textbf{2} & \textbf{3} & \textbf{\(\geq 4\)}   \\ 
        \midrule
        \textbf{1}  & & \( \{D\}, \{c\} \) & \( \{D\}, \{c\} \) & \( \{D\}, \{c\} \) & \( \{D\}, \{c\} \) \\
        \textbf{2}  & & \( \{M,D\}, \{c\} \) & \( \{M,D\}, \{c\} \) & \( \{M,D\}, \{c\} \) & \( \{M,D\}, \{c\} \)  \\
        \textbf{3}  & & \( \{M,D\}, \{c\} \) & \( \{M\}, \{c\} \) & \( \{M\}, \{c\} \) & \( \{M\}, \{c\} \) \\
        \textbf{\(\geq 4\)} & & \( \{M,D\}, \{l, c\} \) & \( \{M\}, \{c\} \) & \( \{M\}, \{c\}  \) & \( \{M\}, \{c\}  \) \\  
        \bottomrule
    \end{tabular}
\end{table}
Similar to our analysis of \(\mathsf{L}\)-rationalizability, along and above the diagonal (i.e.\ where \(n \geq k\)), we obtain the coincidence of \(\mathsf{C}\)-rationalizability with the CH level-\(k\) solution. However, unlike \(\mathsf{L}\)- rationalizability---but similar to downward rationalizability---reasoning of order \(n\) does not imply behavior consistent with reasoning up to order \(n\) in the complete-information game for all sufficiently high-level types (i.e., \(k > n\)). For example, \(\mathsf{C}^3_{2,p,f}(\theta_{2,4}) = \{c\}\), yet \(c \notin R_2^3\).
\end{example}

These observations suggest the following foundation for the CH level-\(k\) solution. The result is formally parallel to the classic level-\(k\) case, but its interpretation is different, because the connection to complete-information rationalizability is weaker.

\begin{theorem}\label{theorem:coincidence-CH}
Let \(p\) be an anchor and \(f\) a level distribution. Then for every \(k \in \mathbb{N}\), every \(n \geq k\), and every \(i \in I\), we have
\[
    \mathsf{C}^n_{i,p,f}(\theta_{i,k}) = CH^k_i[p,f].
\]
\end{theorem}

Thus, for each type \(k\), \(\mathsf{CH}\)-rationalizability reproduces player \(i\)'s classic CH level-\(k\) behavior once the order of higher-order reasoning reaches \(k\). As in the level-\(k\) case, combining the theorem with the epistemic foundation of \(\Delta\)-rationalizability yields an epistemic foundation of the CH model. Unlike in the classic level-\(k\) case, however, this coincidence does not imply the same tight connection to complete-information rationalizability.

\begin{corollary}[Epistemic foundation of the CH model, informal]
Fix a finite order \(n \geq 1\) of higher-order reasoning about rationality, and consider a complete epistemic type structure.\footnote{As in the level-\(k\) case, we require completeness relative to the first-order belief restrictions given by \(\textup{(K1)}\) and \(\textup{(KC)}\), meaning that the type structure contains all belief hierarchies consistent with these restrictions.} Then transparency of \(\textup{(K1)}\), \(\textup{(K2)}\), and \(\textup{(KC)}\), together with rationality and \((n-1)\)-fold belief in rationality, yields behavioral implications coinciding with the classic CH level-\(k\) actions for a type \(k\) with \(k \le n\). Unlike in the classic level-\(k\) case, however, for \(k > n\)\, these behavioral implications need not coincide with the set of \(n\)-rationalizable actions and may even be disjoint from it.
\end{corollary}

\begin{proof}[Proof of \autoref{theorem:coincidence-CH}]
Based on \autoref{CH: bounded_reasoning}, we only have to show that \(\mathsf{C}^k_{i,p,f}(\theta_{i,k}) = CH^k_i[p,f]\) for every \(k \in \mathbb{N}\), which we prove by strong induction.

\noindent \textbf{Base Case (\(k=1\)):} The equality is true, essentially, by definition.

\noindent \textbf{Inductive Step:} Fix \(k \geq 1\) and assume the statement holds for all \(t\) such that \(1 \leq t \leq k\). Now, we prove it holds for \(k+1\).

First, we establish that \(CH^{k+1}_i[p,f] \subseteq \mathsf{C}^k_{i,p,f}(\theta_{i,k})\). For this, consider \(a_i \in CH^{k+1}_i[p,f]\). That is, there exists \(\nu^i \in \Delta (\mathbb{N}_0 \times A_{-i})\) satisfying
\begin{itemize}[nosep]
				\item[(i)] \(a_i \in r_i(\marg_{A_{-i}}\nu^i)\),
                \item[(ii)] \(\marg_{\mathbb{N}_0}\nu^i = f^k\), 
                \item[(iii)] \(\nu^i\left(\left. \cdot \right\vert 0 \right) = p_{-i}\), and 
				\item[(iv)] for \(0 < t \leq k\), \(
                \nu^i\big(CH^t_{-i}[p,f] \big\vert t \big) = 1\).
\end{itemize}
Now, define the following \(\hat G\)-conjecture \(\mu^i\in \Delta(\Theta_{-i} \times A_{-i})\),
\(
    \mu^i(\theta_{-i,t}, a_{-i}) := \nu^i(t, a_{-i}),
\)
for all \((\theta_{-i,t}, a_{-i}) \in \Theta_{-i} \times A_{-i}\). It is clear---using properties (ii) and (iii)---that \(\mu^i\) satisfies the belief restrictions (K1), (K2), and (KC) for type \(\theta_{i,k}\). By properties (ii) and (iv) of \(\nu^i\), we have\footnote{To ease notation, define \(CH_{-i}^0[p,f]:=A_{-i}\).} 
\begin{equation*}
    \mu^i\left(\bigcup_{0 \leq t \leq k} \{\theta_{-i,t}\}\times CH_{-i}^t[p,f] \right)=1.
\end{equation*}
Furthermore, by the inductive hypothesis,
\begin{align*}
      \bigcup_{0 \leq t \leq k} \{\theta_{-i,t}\}\times CH_{-i}^t[p,f] 
      =
      \bigcup_{0 \leq t \leq k} \{\theta_{-i,t}\}\times \mathsf{C}_{-i,p,f}^k(\theta_{-i,t}) 
      \subseteq
      \mathsf{C}_{-i,p,f}^ k.
\end{align*}
Thus, \(\mu^i(\mathsf{C}_{-i,p,f}^ k)=1\). But then, \(a_i \in C^{k+1}_{i,p,f}(\theta_{i,k})\) because property (i) of \(\nu^i\) carries over to \(\mu^i\). 

For the reverse inclusion, consider \(a_i \in \mathsf{C}^{k+1}_{i,p,f}(\theta_{i,k})\). That is, there exists \(\mu^i\in \Delta(\Theta_{-i} \times A_{-i})\) making \(a_i\) a best reply and \(\mu^i(\mathsf{C}_{-i,p,f}^k)=1\). Furthermore, \(\mu^i\) satisfies the belief restrictions (K1), (K2), and (KC) for type \(\theta_{i,k}\). Consider \(\nu^i \in \Delta (\mathbb{N}_0 \times A_{-i})\) defined as
\(
    \nu^i(t, a_{-i}) := \mu^i(\theta_{-i,t}, a_{-i})
\)
for all \((t, a_{-i}) \in \mathbb{N}_0 \times A_{-i}\). By (K1), property (iii) is satisfied for \(\nu^i\). By (KC), property (ii) of \(\nu^i\) holds too. By construction, we also have property (i). It remains to prove that for all \(0 < t \leq k\), \(\nu^i\big(\{t\} \times CH^t_{-i}[p,f] \big) = \nu^i\big(\{t\} \times A_{-i} \big)\).

By the inductive hypothesis and \autoref{CH: bounded_reasoning}, \(CH^t_{-i}[p,f] = \mathsf{C}_{i,p,f}^t(\theta_{i,t})=\mathsf{C}_{i,p,f}^k(\theta_{i,t})\) holds by the inductive hypothesis.
Furthermore,
\(
    \mu^i\left(\{\theta_{-i,t}\} \times \mathsf{C}_{-i,p,f}^k(\theta_{-i,t})\right) = \mu^i\left(\{\theta_{-i,t}\} \times A_{-i}\right) 
\)
holds because \(\mu^i(\mathsf{C}_{-i,p,f}^k)=1\). Thus,
\(
    \nu^i\big(\{t\} \times CH^t_{-i}[p,f] \big) 
    =
    \nu^i\big(\{t\} \times \mathsf{C}_{-i,p,f}^k(\theta_{-i,t}) \big) 
    =
    \nu^i\big(\{t\} \times A_{-i} \big)
\)
\end{proof}

As already noted above, the specific coincidence established in \autoref{theorem:coincidence-CH} does not yield the same close link to complete-information rationalizability as in the level-\(k\) model. Indeed, once the type index and the reasoning order
are allowed to vary independently, consistency with \(n\)-rationalizability can fail much more sharply than under \(\mathsf{L}\)-rationalizability. For some anchors and level distributions, and for every non-trivial order of reasoning, there may be \emph{no type} whose \(\mathsf{C}\)-rationalizable behavior is contained in \(R_i^n\). Thus, unlike in the level-\(k\) case, aggregating behavior across consistent types need not recover the \(n\)-rationalizable actions. This sharper disconnect is one of the main substantive differences between the two models in our framework. The formal definitions, examples, and sufficient conditions are reported in Appendix~\ref{app:subsec:CH_consistency}.

This also makes the robustness question especially natural. If
\(\mathsf{C}\)-rationalizability can depart so substantially from standard \(n\)-rationalizability once anchors and level distributions vary, it becomes important to understand which behavioral implications survive uniformly across those specifications.

\subsection{Robustness}\label{subsec:CH_robust}
Finally, we address a robustness question analogous to those analyzed for our other solution concepts. Specifically, we ask which behavioral implications the assumptions underlying \(\mathsf{C}\)-rationalizability have across all anchors and all level distributions. \autoref{example:CH-rationalization} shows that
\begin{equation*}
    \bigcup_{p \in \Delta(A_1)\times\Delta(A_2), f \in \Delta_+(\mathbb{N}_0)} \mathsf{C}_{i,p,f}^n(\theta_{i,k}) \subseteq R_i^n
\end{equation*}
does not hold in general and therefore illustrate another difference to \(\mathsf{L}\)-rationalizability. On the other hand, by Appendix~\ref{app:subsec:CH_consistency}, in particular \autoref{cor:CH_reasoning_cons}, the inclusion
\begin{equation*}
    \bigcup_{p \in \Delta(A_1)\times\Delta(A_2), f \in \Delta_+(\mathbb{N}_0)} \mathsf{C}_{i,p,f}^n(\theta_{i,k}) \subseteq R_i^1
\end{equation*}
holds trivially for any \(n \in \mathbb{N}\) and any \(k \in \mathbb{N}\). The following example shows that the reverse inclusion need not hold, showing a different answer to the robustness question for downward rationalizability too.
\begin{example}[Justifiable action is not robustly \(\mathsf{C}\)-rationalizable]\label{example:CH-rationalization_inclusion}
Consider the following game.
\begin{table}[ht!]
    \centering
    \caption{A game showing strict inclusion.}
    \label{tab:robust_CH_example}
    \renewcommand{\arraystretch}{0.75} 
    \setlength{\tabcolsep}{10pt} 
    \begin{tabular}{c ccc}
        \toprule
        Player 1 \textbackslash \, 2 & \textbf{\(l\)}  & \textbf{\(c\)} & \textbf{\(r\)}  \\ 
        \midrule
        \textbf{\(U\)}  & \(3, 2\) & \(0, 0\) & \(0, 1\)  \\
        \textbf{\(M\)}  & \(0, 0\) & \(3, 2\) & \(0, 1\)  \\
        \textbf{\(D\)}  & \(9, 2\) & \(9, 2\) & \(9, 0\)  \\ 
        \bottomrule
    \end{tabular}
\end{table}

In this game, all actions of player 2 are justifiable. That is, \(R_2^1 = \{l,c,r\} = A_2\).

However, we have the following claim.

\begin{claim}
  For every level distribution \(f \in \Delta_+(\mathbb{N})\), every anchor \(p \in \Delta(A_1)\times\Delta(A_2)\), and every \(k,n \geq2\), 
  \(
    r \notin \mathsf{C}_{2,p,f}^n(\theta_{2,k}).
  \)
\end{claim}

\begin{proof}
Throughout, fix an arbitrary level distribution \(f \in \Delta_+(\mathbb{N})\).

Note that player 1 has a dominant action, \(D\). Thus for every anchor and every \(k,n\in\mathbb{N}\), \(\mathsf{C}_{1,p,f}^n(\theta_{1,k}) = \{D\}\).

For player 2, \(r\) is a best reply only to the \(G\)-conjecture that is the fifty-fifty mixture between \(U\) and \(M\).

First, if \(p_1\) is not this mixture, then \(r\notin \mathsf{C}_{2,p,f}^n(\theta_{2,k})\) for all \(k,n\in\mathbb{N}\).

Second, if \(p_1\) is this mixture, then \(r\in \mathsf{C}_{2,p,f}^1(\theta_{2,k})\) for any \(k \geq 1\). However for any order of reasoning \(n \geq 2\), by (KC), each type \(\theta_{2,k}\) with \(k\geq 2\) must assign strictly positive probability to elements of
\(
  \bigcup_{t=1}^{k-1} \mathsf{C}_{1,p,f}^{n-1}(\theta_{1,t}) = \{D\}.
\) Hence any \(\hat G\)-conjecture \(\mu^2\) satisfying the belief restrictions must satisfy \(\marg_{A_1}\mu^2(D)>0\), so \(r\) cannot be justified.
\end{proof}
\end{example}

The previous example crucially relies on a non-generic game \(G\). Indeed, if we restrict attention to generic games, we do get the non-trivial inclusion too. The key argument is a generalization of the idea in \autoref{example:CH-rationalization_running}.
\begin{proposition}\label{prop:generic_CH_justfiable}
Let \(G\) be generic,\footnote{We say a game is \emph{generic} if, for every player and every justifiable action, there exists a \(G\)-conjecture making that action the unique best reply.} then for every \(i\in I\), every \(n\in\mathbb{N}\), and every \(k\in\mathbb{N}\),
\begin{equation*}
\bigcup_{p \in \Delta(A_1)\times\Delta(A_2), f \in \Delta_+(\mathbb{N}_0)} \mathsf{C}_{i,p,f}^n(\theta_{i,k}) = R_i^1.
\end{equation*}
\end{proposition}

\begin{proof}
    We only prove the remaining, non-trivial, inclusion. Thus, fix a player \(i \in I\), and an action \(a_i \in R_i^1\). By genericity, there is a \(G\)-conjecture \(\nu^i\) making \(a_i\) the unique best reply. Furthermore, since the game is finite, there is an open ball around \(\nu^i\) such that uniqueness holds for all the conjectures in this ball. Call this ball \(B_i\).

    Now, set \(p_{-i}=\mu^i\) and consider the following class of level distribution indexed by \(\varepsilon \in (0,1/2)\):
   \(
    f_\varepsilon(0) = \frac{1-2\varepsilon}{1-\varepsilon}, \text{ and for every } k\geq 1, f_\varepsilon(k)=\varepsilon^k.
    \)
    Note that there exists \(\bar \varepsilon \in (0,1/2)\) such that a \(\hat{G}\)-conjecture \(\mu^i\) satisfying the belief restrictions with level distribution \(f_\varepsilon\) and \(\varepsilon \leq \bar \varepsilon\) will have that \(\marg_{A_i} \mu^i \in B_i\). Thus, \(a_i \in \mathsf{C}_{i,p,f_\varepsilon}^n(\theta_{i,k}) \) holds for all \(k,n \in \mathbb{N}\).
\end{proof}

	\bibliographystyle{ecta}
	\bibliography{ref}

\newpage
\appendix
\section{Consistency Across Reasoning and the Connection to BFK}\label{app:sec:consistency}

This appendix collects the consistency-across-reasoning results for \(\mathsf{L}\)- and \(\mathsf{C}\)-rationalizability and expands on the relation to \citet{brandenburger2020two}. The purpose is twofold. First, it records the formal statements behind the summary discussion in Sections~5 and~6. Second, it makes explicit how our payoff-uncertainty approach relates to the complete-information epistemic construction of \citet{brandenburger2020two}.

\subsection[L-Rationalizability: Consistency Across Reasoning]{\(\mathsf{L}\)-Rationalizability: Consistency Across Reasoning}\label{app:subsec:Lk_consistency}
Interestingly, in \autoref{example:L-rationalization}, we observe \(k\)-types reasoning to a higher order \(n > k\) whose behavior is inconsistent with rationality and \(n\)-th order belief in rationality in the complete-information game \(G\). In particular, in this example, we have \(\mathsf{L}_{1,p}^2(\theta_{1,1})=\{D\} \not \subseteq R_1^2 = \{U,M\}\).

On the other hand, if we fix a given order of reasoning \(n\) and consider only those \(k\)-types that are consistent with strategic reasoning up to order \(n\) in the complete-information game, we obtain exactly the same actions as through the usual \(n\) iterations of rationalizability. More formally, for a given order of \emph{reasoning} \(n \in \mathbb{N}\), define
\[
K_{i,p}^n := \big\{k \in \mathbb{N} \big\vert \mathsf{L}^n_{i, p}(\theta_{i,k}) \subseteq \rat_i^n\big\}
\]
as the set of player \(i\)'s type indices corresponding to \(k\)-types that are consistent with strategic reasoning up to order \(n\) in the complete-information game. Trivially,
\[
\bigcup_{k \in K^n_{i,p}} \mathsf{L}^n_{i, p}(\theta_{i,k}) \subseteq \rat_i^n
\]
for every \(i \in I\) and every \(n \in \mathbb{N}_0\). The more interesting observation is that the reverse inclusion also holds. Furthermore, for every order of reasoning \(n\), there always exists at least one consistent type---specifically, the type whose level matches the reasoning order. Thus, \(\mathsf{L}\)-rationalizability respects rationalizability in the complete-information game.

\begin{corollary}\label{corollary:bfk_comparison}
For every anchor \(p\), every \(n \in \mathbb{N}_0\), and every \(i \in I\),
\begin{enumerate}[nosep]
    \item \(\bigcup_{k \in K^n_{i,p}} \mathsf{L}^n_{i, p}(\theta_{i,k}) = \rat_i^n\), and
    \item \(n \in K^n_{i,p}\) (\emph{a fortiori}, \(\lk[i]^n \subseteq \rat_i^n\)).    
\end{enumerate}
\end{corollary}

\begin{proof}
Fix an anchor \(p\), and consider an arbitrary \(n \in \mathbb{N}_0\) and \(i \in I\).
\begin{enumerate}[nosep]
    \item This follows directly from \autoref{prop:classic_k_respects_rat}, in particular, that \(\mathsf{L}^n_{i, p}(n_{i}) \subseteq R^n_i\). The second part then follows from \autoref{theorem:coincidence-classic}.
    \item For the non-trivial inclusion, observe that by \autoref{theorem:coincidence-classic}, for any \(k > n\), we have \(\mathsf{L}^n_{i, p}(\theta_{i,k}) = \rat_i^{n}\), implying that any such \(k\) belongs to \(K^n_{i,p}\). The conclusion follows.
\end{enumerate}
\end{proof}

\subsection[Detailed Comparison With BFK]{Detailed Comparison With \citet*{brandenburger2020two}}\label{app:subsec:BFK}

Similar to our analysis in the previous section, \citet{brandenburger2020two} study the classic level-\(k\) solution from an epistemic perspective. As in our analysis, they focus their investigation on complete information games per se. However, in crucial contrast to our approach of embedding the complete-information game into an incomplete-information one, they retain the original game and analyze a corresponding \emph{epistemic game}. Specifically, for a given anchor \(p\), they append an \(A\)-based (epistemic) level-\(k\) type structure 
\(
  \mathcal{T} = \bigl(A_{-i},\,T_i,\,\beta_i\bigr)_{i\in I}.
\)
Given this type structure, for each \(i\in I\) they consider a Borel cover \((T_i^k)_{k=1}^\infty\) of the type space \(T_i\) satisfying the following conditions:\footnote{These two conditions correspond to (i) and (ii) in their Definition 4.2, respectively.}
\begin{enumerate}[leftmargin=2.5\parindent, align=left, nosep]
    \item[\textbf{BFK-1}] For each \(t_i \in T_i^1\), \(\marg_{A_i}\beta_i(t_i) = p_{-i}\), and
    \item[\textbf{BFK-L}] For each \(k\) and each \(t_i \in T_i^{k+1}\), \(\beta_i(t_i)(A_{-i} \times T_{-i}^{k}) = 1\).
\end{enumerate}

Within this framework, their epistemic condition is rationality and \((k-1)\)-th order belief in rationality (R\((k-1)\)BR). If the type structure is rich enough, the behavioral consequence of this assumption is characterized by \(\lk[i]^k\) for any epistemic type in \(T_i^k\) (Theorem 6.1 (ii)) and by \(\rat_i^k\) across all epistemic types (Theorem 5.1 (ii)).

At first glance, \citeauthor{brandenburger2020two}'s model appears similar to ours. Indeed, BFK-1 and BFK-L closely resemble our (K1) and (KL) conditions, respectively. Moreover, \citeauthor{brandenburger2020two}'s (\citeyear{brandenburger2020two}) Theorem 6.1 (ii) and our \autoref{theorem:coincidence-classic} both characterize the classic level-\(k\) solution, and elements of their proof of Theorem 5.1 (ii) can be traced in our proofs of \autoref{theorem:coincidence-classic} and \autoref{corollary: indifference}.

Despite these similarities, the two models differ in fundamental ways. In their approach, the complete information game is extended with \emph{epistemic} types, and, as is standard in epistemic game theory, epistemic assumptions impose restrictions on behavior by eliminating certain types. Specifically, R\((k-1)\)BR serves to exclude epistemic types that satisfy \((k-1)\)-th order belief in rationality but do not satisfy \(k\)-th order belief.

In contrast, our model employs \emph{belief (i.e., level-\(k\)) types} within a game with \emph{payoff uncertainty}. Since \(0\)-types have a constant payoff function, their (randomized) behavior described by \(p_i\) is rationalizable. Consequently, our conditions (K1) and (KL) do not lead to the elimination of any \(k\)-types during the iterative procedure. In other words, our framework incorporates level-\(k\) reasoning directly into the classic rationalizability argument formalized using epistemic game theory, without requiring players to rule out level-\(k\) types beyond what the classic level-\(k\) motivation assumes.


Indeed, our model can accommodate the additional requirement that level-\(k\) types behave according to the usual strategic reasoning in the complete information game up to some order \(n\). We formalized this notion earlier with \(K^n_{i,p}\). With this in mind, \autoref{corollary:bfk_comparison}(1) closely aligns with Theorem 5.1 (ii) of \citet{brandenburger2020two}, while \autoref{corollary:bfk_comparison}(2) corresponds to Theorem 6.1 (ii) of \citet{brandenburger2020two}.

\subsection[C-Rationalizability: Consistency Across Reasoning]{\(\mathsf{C}\)-Rationalizability: Consistency Across Reasoning}\label{app:subsec:CH_consistency}
Next, we explore the consistency of behavior across reasoning in the incomplete-information game and the complete-information game. Similar to before, for a given reasoning order \(n \in \mathbb{N}\), define
\[
K_{i,p,f}^n := \{k \in \mathbb{N}\mid \mathsf{C}_{i,p,f}^n(\theta_{i,k}) \subseteq R_i^n\}.
\]
This set identifies all types whose behavior at reasoning order \(n\) is consistent with the predictions implied by rationality and higher-order belief in rationality in the complete-information game.

Here, the results differ sharply from those for \(\mathsf{L}\)-rationalizability, revealing a fundamental distinction between the two seminal level-\(k\) approaches. While \(\mathsf{L}\)-rationalizability maintains a close alignment with rationalizability in the complete-information game, no such connection holds for \(\mathsf{C}\)-rationalizability.

\begin{corollary}\label{cor:CH_reasoning_cons}
Let \(p\) be any anchor, \(f\) any level distribution, and \(i\in I\). Then,
\begin{enumerate}
  \item \(K_{i,p,f}^1 = \mathbb{N}\); and
  \item for every \(n \in \mathbb{N}\), \(\displaystyle\bigcup_{k\in K_{i,p,f}^n} \mathsf{C}_i^n(\theta_{i,k}) \subseteq R_i^n.\)
\end{enumerate}
\end{corollary}

\begin{proof}
\begin{enumerate}
  \item By construction, only justifiable actions can ever appear in the CH level-\(k\) process.
  \item This inclusion follows immediately from the definition of \(K_{i,p,f}^n\).
\end{enumerate}
\end{proof}

Although \autoref{cor:CH_reasoning_cons} is straightforward, it is the strongest general statement that can be made. It implies that consistency with the behavioral implication of rationality in the complete-information game always holds, but that consistency may fail at any non-trivial order of reasoning. Indeed, the following example demonstrates the tightness of this result. There exists a game, an anchor \(p\), and a level distribution \(f\) such that for some player \(i\in I\) and every \(n>1\),
\begin{enumerate}
  \item \(K_{i,p,f}^n = \emptyset\); and
  \item \(\displaystyle\bigcup_{k\in K_{i,p,f}^n} \mathsf{C}_{i,p,f}^n(\theta_{i,k}) \subsetneq R_i^n.\)
\end{enumerate}
In this case, no type remains consistent with higher-order reasoning, and therefore the set of actions supported by \(\mathsf{C}\)-rationalizability across consistent types becomes strictly smaller than the set of \(n\)-rationalizable actions. The emptiness of \(K_{i,p,f}^n\) suggests an \emph{epistemic impossibility} result, although we do not pursue that interpretation formally here.\footnote{In examples where \(K_{i,p,f}^n=\emptyset\), one would expect the corresponding epistemic event encoding \(n\)-th order belief in rationality in the complete-information game, \'{a} la \citet{brandenburger2020two}, to be empty once it is combined with the reasoning implicit in the CH model. We conjecture that this statement can be formalized using the tools of \citet{brandenburger2020two}.}

\begin{example}[continued from \autoref{example: downward-rationalization}]\label{example:CH-rationalization_running}
The game in our running example (\autoref{tab:iterated_elimination_example}) is again illustrative. Recall that \(D \notin R_1^2\). For \(\mathsf{C}\)-rationalizability, consider the anchor \(p = (\delta_D,\delta_r)\) and a level distribution \(f\) that intuitively reflects a lexicographic ordering of levels. Namely, level-\(0\) is infinitely more likely than level-\(1\), which in turn is infinitely more likely than level-\(2\), and so on. Formally, fix \(\varepsilon \in (0,\tfrac14)\), and define
\begin{equation*}
    f(0) = \frac{1-2\varepsilon}{1-\varepsilon}, \text{ and for every } k\geq 1, f(k)=\varepsilon^k.
\end{equation*}

\begin{claim}
    For every \(k,n \in \mathbb{N}\), \(C \in \mathsf{C}_1^n(k_1)\).
\end{claim}
\begin{proof}
    Note that for any \(k_1\), any \(\hat G\)-conjecture satisfying (KC) must assign probability strictly greater than \(\tfrac{1 - \tfrac12}{1 - \tfrac14} = \tfrac23\) to level-0 of the co-player. Thus, by (K1), any such conjecture must also assign probability strictly greater than \(\tfrac23\) to the anchor, \(r\). Under any such conjecture, \(D\) yields an expected utility strictly greater than \(\tfrac73\). Conversely, \(U\) and \(M\) yield expected utilities strictly less than \(\tfrac53\) and \(\tfrac73\), respectively.
\end{proof}
Hence, for every \(n \geq 2\),
\(
    \mathsf{C}_{1,p,f}^n(\theta_{1,k}) \not\subseteq R_1^n,
\)
so \(K_{1,p,f}^n = \emptyset\). Consequently, for every \(n \geq 2\),
\(
    \bigcup_{k \in K_{1,p,f}^n} \mathsf{C}_{1,p,f}^n(\theta_{1,k}) = \emptyset \subsetneq R_1^n,
\)
because \(R_1^n \neq \emptyset\) for every \(n \in \mathbb{N}_0\).
\end{example}

A sufficient condition for ensuring that, for every order of reasoning \(n\), there exists at least one level-\(k\) type consistent with that level of reasoning in the complete-information game is that the anchor is supported on the set of rationalizable action profiles.

\begin{corollary}\label{cor:CH_anchor_support}
Let \(p\) be any anchor, \(f\) any level distribution, \(n \in \mathbb{N}_0\), and \(i \in I\). If, for every \(j \in I\), \(p_j\bigl(R_j^\infty\bigr) = 1\), then \(K_{i,p,f}^n \neq \emptyset\).
\end{corollary}

\begin{proof}
By the best-reply property\footnote{See, for example, \citet[Chapter 4.3]{battigalli2023game}.} of \(R^\infty\), for every \(k \geq 1\) we have
\(
    \mathsf{C}_{i,p,f}^k(\theta_{i,k}) \subseteq R_i^\infty \subseteq R_i^k.
\)
Hence, \(k\in K_{i,p,f}^n\), so \(K_{i,p,f}^n\neq\emptyset\).
\end{proof}

Using similar wording as \citet[p.14]{crawford2013structural}, the proof of \autoref{cor:CH_anchor_support} reveals that the CH level-\(k\) solution respects \(k\)-rationalizability if the anchor is concentrated on the set of rationalizable action profiles. Clearly, however, this condition is sufficient but not necessary.\footnote{For example, in any game in which both players have a dominant action, one obtains \(\theta_{i,k}^n = \mathbb{N}\neq\emptyset\) for all \(i\in I\) and every level distribution \(f\), even if \(\supp p\not\subseteq R_i^\infty\).}

\end{document}